\documentclass[11pt,a4paper]{article}
\usepackage{jheppub,xcolor}
\pdfoutput=1
\usepackage[utf8]{inputenc}
\graphicspath{{./figs/}}

\usepackage{amssymb}
\usepackage{dcolumn}	
\usepackage{bm}			
\usepackage{bbm} 		
\usepackage{multirow}
\usepackage{slashed}
\usepackage{blkarray}
\usepackage{booktabs}
\usepackage{makecell}

\definecolor{Red}{rgb}{1.,0.,0.}

\newcommand{\mcM}{\mathcal{M}}
\newcommand{\mcO}{\mathcal{O}}

\newcommand{\rtsha}{\sqrt{\hat{s}}}
\newcommand{\mw}{M_W}
\newcommand{\mh}{M_h}

\newcommand{\Oqtrip}{\mcO_{\varphi q}^{(3)}}
\newcommand{\Ohw}{\mcO_{\varphi {\textsc w}}}
\newcommand{\Ohwtil}{\mcO_{\varphi \widetilde {\textsc w}}}
\newcommand{\Cqtrip}{c_{\varphi q}^{(3)}}
\newcommand{\Chw}{c_{\varphi {\textsc w}}}
\newcommand{\Chwtil}{c_{\varphi \widetilde {\textsc w}}}

\newcommand{\bc}{\begin{center}}
\newcommand{\ec}{\end{center}}
\newcommand{\ba}{\begin{array}}
\newcommand{\ea}{\end{array}}

\newcommand{\pth}{p_{T}^{h}}

\def\Re{{\rm Re\,}}

\preprint{DESY 20-054, ZU-TH 09/20, PSI-PR-20-16}

\title{A New Precision Process at FCC-hh: the  diphoton leptonic Wh channel}

\author[a]{Fady Bishara,}
\author[a,b]{Philipp Englert,}
\author[a,b]{Christophe Grojean,}
\author[a,c,d]{Marc Montull,}
\author[e]{Giuliano Panico}
\author[a, b]{and Alejo N. Rossia}
\affiliation[a]{Deutsches Elektronen-Synchrotron (DESY), D-22607 Hamburg, Germany}
\affiliation[b]{Institut f{\"u}r Physik, Humboldt-Universit{\"a}t zu Berlin, D-12489 Berlin, Germany} 
\affiliation[c]{Paul  Scherrer  Institut,  CH-5232  Villigen  PSI,  Switzerland}
\affiliation[d]{Physik-Institut,  Universit{\"a}t  Z{\"u}rich,  Winterthurerstrasse  190,  CH-8057  Z{\"u}rich,  Switzerland}
\affiliation[e]{Universit\`{a} di Firenze, Via Sansone 1, 50019 Sesto Fiorentino, Florence, Italy}

\emailAdd{fady.bishara@desy.de}
\emailAdd{philipp.englert@desy.de}
\emailAdd{christophe.grojean@desy.de}
\emailAdd{marc.montull@gmail.com}
\emailAdd{giuliano.panico@unifi.it}
\emailAdd{alejo.rossia@desy.de}

\date{\today}
\abstract{The increase in luminosity and center of mass energy at the FCC-hh will open up new clean channels where BSM contributions are enhanced at high energy. In this paper we study one such channel, $Wh \to \ell\nu\gamma\gamma$. We estimate the sensitivity to the $\Oqtrip, \, \Ohw$, and $\Ohwtil$ SMEFT operators. We find that this channel will be competitive with fully leptonic $WZ$ production in setting bounds on $\Oqtrip$. We also find that the double differential distribution in the $p_T^h$ and the leptonic azimuthal angle can be exploited to enhance the sensitivity to $\Ohwtil$. 
However, the bounds on $\Ohw$ and $\Ohwtil$ we obtain in our analysis, though complementary and more direct, are not competitive with those coming from other measurements such as EDMs and inclusive Higgs measurements.
}

\keywords{}

\begin{document}
\maketitle
\flushbottom

\section{Introduction}
\label{sec.intro}

Hadron colliders are typically perceived as wonderful machines for direct searches of new particles, but with a limited impact on precision measurements, especially regarding electroweak (EW) observables. There are, however, noteworthy exceptions to this statement. Thanks to the interplay between the large accessible energy
range and a clever selection of channels with relatively low uncertainties, one can perform precision measurements at hadron colliders that can compete with the ones possible at lepton machines such as LEP~\cite{Farina:2016rws,deBlas:2013qqa}. 

The key ingredient that allows for this enhanced precision is that
new physics effects tend to grow with the center of mass energy.
If we parametrize the deviations within an effective field theory (EFT) formalism, the leading SM deformations, which generically correspond to operators of dimension six, give rise to amplitudes that can grow up to quadratically with the energy of the process.
In such a situation, having access to the high-energy tails of the kinematic distributions can significantly enhance the achievable precision.

It has been shown that, at the LHC, several simple two-body production channels can be exploited to obtain precision measurements~\cite{Farina:2016rws,deBlas:2013qqa,Domenech:2012ai,Farina:2018lqo,Biekotter:2018rhp,Baglio:2020oqu,Alioli:2020kez,Boughezal:2020uwq}. Among them, diboson production processes, featuring EW gauge bosons or the Higgs boson,
play a privileged role since they can be used to indirectly test the high-energy Higgs dynamics~\cite{Biekotter:2018rhp,Baglio:2020oqu,Falkowski:2015jaa,Butter:2016cvz,Azatov:2017kzw,Franceschini:2017xkh,Liu:2019vid,Bellazzini:2018paj,Banerjee:2018bio,Grojean:2018dqj,Baglio:2018bkm,Almeida:2018cld,Azatov:2019xxn,Banerjee:2019pks,deBlas:2019wgy,Brehmer:2019gmn,Henning:2019vjr,Chiu:2019ksm,Baglio:2019uty,Banerjee:2019twi,Freitas:2019hbk}. 

\begin{figure}[t]
    \centering
    \includegraphics{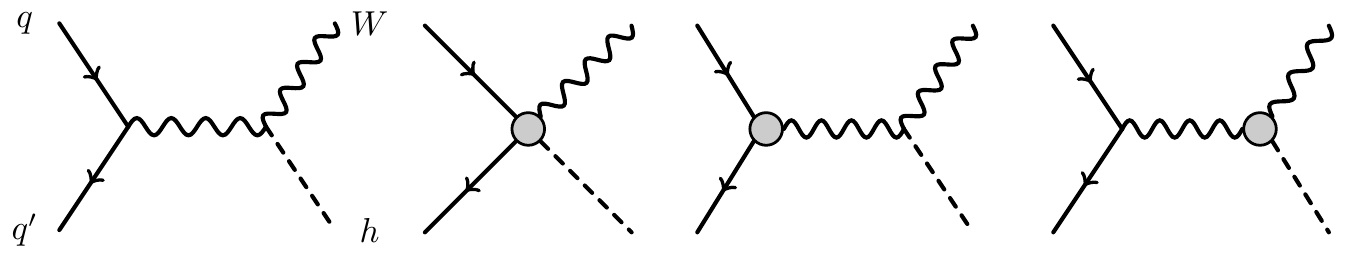}
    \caption{Representative Feynman diagrams for $q\,q'\to Wh$ at leading order. The leftmost diagram shows the SM process while the gray circles in the other diagrams represent one insertion of a dimension-6 operator.}
    \label{fig:feynman}
\end{figure}

In this paper, we will focus on a specific diboson channel, $Wh$, where the $W$ decays leptonically. Figure~\ref{fig:feynman} shows the leading order SM Feynman diagram (leftmost diagram). At the LHC, this channel can be exploited~\cite{Butterworth:2015bya,Franceschini:2017xkh,Liu:2019vid,Banerjee:2019twi} for precision measurements by only considering decays of the Higgs into a pair of bottom quarks, especially thanks to jet substructure analysis~\cite{Butterworth:2008iy}.
To give an idea of how this situation will change when the next generation colliders are ready, we show in Table~\ref{tab:events_comparison} the approximate number of $Wh$ events expected for different Higgs decay channels at the LHC and future hadron colliders. These results correspond to the leading order SM prediction for the number of events with high Higgs transverse momentum ($p_T^{h} > 550\, {\rm GeV}$). The $W$ is assumed to decay to first and second generation leptons and only detector acceptance cuts were applied (see upper part of Table~\ref{tab:gen_cuts}). We considered three benchmark colliders: the high-luminosity LHC (HL-LHC), at $14\,$TeV and $3\,{\rm ab}^{-1}$,
the high-energy LHC (HE-LHC), at $27\,$TeV and $15\,{\rm ab}^{-1}$, and the FCC-hh at $100\,$TeV and $30\,{\rm ab}^{-1}$.  

One can see that rare channels, such as the final states with the Higgs decaying into two photons or two muons, have branching ratios that are too small to populate the high-energy tail at HL-LHC.
At future high-energy colliders, the situation will improve drastically thanks to a big increase in the production cross section ($\sim 30\times$) and the possibility to collect significantly more integrated luminosity ($\sim 10\times$).
For instance, at FCC-hh, the $\gamma \gamma$ channel is expected to provide $\sim 700$ events, which can allow one to probe new physics effects at the $5$--$10\%$ level.

A clear advantage of these rare decay channels is the fact that the final-state configuration can be easily reconstructed and  background processes are small. In such cases, very simple analysis strategies can give competitive results.
In this paper, we study the Higgs to two photon channel at the FCC-hh. 
The complementary $Z(h\to\gamma\gamma)$ channel with $Z\to\ell\ell$ and $Z\to\nu\nu$ also becomes accessible at the FCC-hh but we leave its investigation for future work.

\begin{table}[t]
\centering
$\begin{array}{c c c c c}
\hline
\text{Higgs decay} \quad \quad &  \text{Higgs BR} \quad \quad & n_\textsc{hl-lhc}  \quad \quad & n_\textsc{he-lhc}  \quad \quad & n_{\textsc{fcc}\textrm{\tiny -hh}}  \quad \quad  \\
\hline
\vrule height 13pt depth 5pt width 0pt
\bar{b} b		& 0.6			& 600 		& 1\cdot10^4 	& 2 \cdot 10^5 \\
\tau \tau 		&  6\cdot10^{-2}			&  60 		&  1 \cdot 10^3 	& 2 \cdot 10^4 \\
\gamma \gamma	& 2 \cdot 10^{-3} 	& 2		& 40	& 700\\
\mu \mu 		& 2\cdot 10^{-4} 	& 0.2 	& 4		& 70 \\
4 \ell			& 1\cdot 10^{-4} 	& 0.1	& 2		& 40 \\
\hline
\end{array}$
\caption{Number of $Wh \to \ell \nu \, XX$ events predicted by the SM at LO for different Higgs decay channels and with
a cut $p_T^{h} > 550\,{\rm GeV}$. The results correspond to $3 \, \text{ab}^{-1}$, $|\eta|<2.5$ for the HL-LHC, $15 \, \text{ab}^{-1}$, $|\eta|<6$ for the HE-LHC and $30 \, \text{ab}^{-1}$, $|\eta|<6$ for the FCC-hh.}
\label{tab:events_comparison}
\end{table}

\medskip

The paper is structured as follows. In Section~\ref{sec:EFT}, we discuss the general features
of the $Wh$ production channel and the main new physics effects that can be tested through
its study. Also in Section~\ref{sec:EFT}, we estimate the expected size of the dimension-six Wilson coefficients in generic BSM scenarios.
In Section~\ref{sec:signal_background}, the details of our analysis are presented. In particular, we discuss the features of the signal and background processes and the cut-flow we devised to enhance the sensitivity to new physics effects.
The results of the analysis are collected in Section~\ref{sec.results}, while the summary of our work and some future directions are discussed in Section~\ref{sec:conclusions}.
Finally, we collect in Appendices~\ref{app:HelAmps}, \ref{app:mc_evt_gen}, and~\ref{app:other} some additional details that were not included in the main text.

\section{Theoretical background}
\label{sec:EFT}

\subsection{High energy sensitivity and interference patterns}\label{sec:interference}

In order to parametrize new physics effects we adopt the EFT formalism, focusing on the leading SM deformations
corresponding to dimension-6 operators. We restrict our attention to operators that induce a growth in the $Wh$ amplitude in the high energy limit. We further assume that new physics obeys the minimal flavor violation hypothesis~\cite{DAmbrosio:2002vsn,Chivukula:1987py,Hall:1990ac,Buras:2003jf,Buras:2000dm}.
Hence, we neglected dipole operators or those generated by right-handed charged currents, since they are suppressed by light Yukawa couplings (see Refs~\cite{Alioli:2017ces,Alioli:2018ljm} for a study of scenarios where right-handed charged currents are not subject to MFV suppression). 
In the Warsaw basis~\cite{Grzadkowski:2010es}, we can therefore restrict our attention to the three operators:
\begin{eqnarray}
{\cal O}_{\varphi q}^{(3)} &=&\left(\overline{Q}_{L} \sigma^{a} \gamma^{\mu} Q_{L}\right)\left(i H^{\dagger} \sigma^{a} \stackrel{\leftrightarrow}{D}_{\mu} H\right)\,, \\
\Ohw &=& H^\dagger H\, W^{a, {\mu\nu}} W^a_{\mu\nu}\,, \\
\Ohwtil &=& H^\dagger H\, W^{a, {\mu\nu}} \widetilde {\textsc W}^a_{\mu\nu}\,.
\end{eqnarray}
where $\sigma^a$ are the Pauli matrices and $\widetilde W^{a, {\mu\nu}} \equiv 1/2\, \epsilon^{\mu\nu\rho\sigma} W^a_{\rho \sigma}$. We define dimensionless Wilson coefficients for the effective operators by introducing explicit powers of the cutoff $\Lambda$. For instance, the coefficient of $\Oqtrip$ is $c_{\varphi q}^{(3)}/\Lambda^2$; analogous conventions are used for the other operators.

In our analysis we neglect any modification of the Higgs branching ratio to
$\gamma \gamma$.\footnote{Notice that since $\Ohw$ and $\Ohwtil$ modify the $H \to \gamma\gamma$ branching ratio, for large values of the corresponding Wilson coefficients, some cancellation must take place. For instance additional contributions coming from ${\cal O}_{\varphi B}$ could provide such a cancellation, as happens in minimally coupled models.
}

This is justified because by the end of the HL-LHC the bound on the effective $h \gamma \gamma$ coupling is expected to be below $2 \%$, and below $0.3 \%$ after FCC-ee+FCC-hh, from a global analysis of Higgs data~\cite{deBlas:2019rxi}.
Furthermore, the CP-odd operator, $\Ohwtil$, is strongly constrained by EDM measurements~\cite{Dekens:2013zca,Panico:2018hal,Cirigliano:2019vfc} as we discuss in Section~\ref{sec.results}.

In order to maximize the sensitivity of our analysis to BSM effects, 
it is useful to analyze the interference between the SM and the new physics contributions. We recall here that the presence of interference between the SM and the BSM contributions is a key ingredient to enhance the
sensitivity, since in our case of study the SM term always dominates over the BSM contribution. Indeed, in the absence of interference, the BSM contributions come
uniquely from the square of the new physics amplitude and become visible only for very large values of the Wilson coefficients.\footnote{Such a situation is also problematic because additional contributions from dimension-8 operators, which we do not include in our analysis, could play an important role, making the bounds more model dependent~\cite{Contino:2016jqw}.}
\begin{table}[t]
\begin{center}
\begin{tabular}{c | cccc}
  \toprule
  $W$ polarization & SM & ${\cal O}_{\varphi q}^{(3)} $ & ${\cal O}_{\varphi {\textsc w}}$ & ${\cal O}_{\varphi \widetilde{\textsc w}}$ \\
  \midrule
  $\lambda=0$ & $1$  &	 $\dfrac{\hat{s}}{\Lambda^2}$  & $\dfrac{M_W^2}{\Lambda^2}$ & 0 \\[4mm]
  $\lambda=\pm$ &$ \dfrac{M_W}{\sqrt{\hat{s}}} $ & $\dfrac{\sqrt{\hat{s}}\, M_W}{\Lambda^2}$ & $\dfrac{\sqrt{\hat{s}}\, M_W}{\Lambda^2}$ &
  $\dfrac{\sqrt{\hat{s}}\,M_W}{\Lambda^2}$\\
  \bottomrule
\end{tabular}
\end{center}
\caption{High energy behavior of the SM and BSM helicity amplitudes for $pp\rightarrow Wh$.
}
\label{tab:Operator_growth}
\end{table}

The leading high-energy behavior of each helicity amplitude is shown in Table~\ref{tab:Operator_growth}. The leading SM amplitude is the one with a longitudinally-polarized $W$ boson, which behaves as a constant at high center-of-mass energy of the process, $\sqrt{\hat s}$, whereas the transverse polarization channels are suppressed at high energy. This contrasts with the case of $WZ$ production, where the $(+,-)$ and $(-,+)$ polarization channels, which are obviously absent for $Wh$, have the same energy behaviour as the longitudinal one and constitute the main source of background for the longitudinal BSM signal at high energy (at least for leptonic $WZ$ decays).
The only new physics operator that induces a growth of order $\hat s/\Lambda^2$ in the $Wh$ amplitude is ${\cal O}_{\varphi q}^{(3)}$, while $\Ohw$ and $\Ohwtil$ generate amplitudes that grow at most with $\sqrt{\hat s}/\Lambda$. It is interesting to notice that the $\Oqtrip$ operator mainly contributes to the longitudinal $W$ channel
and therefore can lead to a strong interference with the SM amplitude.
On the contrary, the leading contributions from the $\Ohw$ and $\Ohwtil$ operators
are in the transverse $W$ channels, which are subleading for the SM.

\subsubsection*{Differential analysis in $p_T^h$}
If one performs an analysis by integrating over the $W$ decay products,
only amplitudes with the same $W$ polarizations can interfere with each other in the high energy limit. In this case the SM squared amplitude and the leading interference terms have the following $\sqrt{\hat s}$ and $\theta$ dependence
\begin{eqnarray}
\left| \mcM_{SM} \right|^2 &\sim& \sin^2 \theta \,, \hspace{1.6cm} \Re\mcM_{SM} \, \mcM_{\varphi {\textsc w}}^* \sim \frac{M_W^2}{\Lambda^2}\,, \nonumber \\
\Re\mcM_{SM} \, \mcM_{\varphi q}^{(3) \, *} &\sim& \frac{\hat s}{\Lambda^2} \sin^2 \theta \,,
 \hspace{1.1cm} \Re\mcM_{SM} \, \mcM_{\varphi \widetilde {\textsc w}}^* =0 \,,
 \label{eq:inclusiveInterf}
\end{eqnarray}
where $\theta$ is the scattering angle of the $W$ boson (see Fig.~\ref{fig:decay_angles}). We provide the full expressions for the helicity amplitudes in Appendix~\ref{app:amps_wh}.  

The interference term between $\Ohw$ and the SM is constant and no enhancement with respect to the SM amplitude is present, while the interference between ${\cal O}_{\varphi q}^{(3)}$ and the SM goes like $\hat s/\Lambda^2$. Finally, the $\Ohwtil$ amplitude does not interfere with the SM amplitude because it is CP odd and can only enter quadratically in the cross section.

An inclusive analysis in the $W$ decay products and differential in $p_T^h$ (which is correlated with $\sqrt{\hat s}$) is therefore expected to provide a good sensitivity to the $\mathcal{O}_{\varphi q}^{(3)}$ operator, but to be rather inefficient in
probing $\Ohw$ and $\Ohwtil$. 

\subsubsection*{Double differential analysis in $p_T^h$ and $\phi_W$}

If one considers differential distributions in the decay
angles of the $W$ boson products, the interference between different helicity channels (and CP parities) can be restored~\cite{Panico:2017frx,Azatov:2017kzw,Azatov:2019xxn,Banerjee:2019pks}. 
This can be checked explicitly by computing the fully differential amplitudes for $p p  \to W h \to \ell \nu h$ and looking at the interference terms.
For reference, we write down, schematically, the behavior of the squared SM amplitude and the interference terms, see Appendix \ref{app:amps_squared} for the full expressions.
The leading terms in the $\mw/\sqrt{\hat s}$ expansion that depend on the $W \to \ell \nu$ decay angles are
\begin{equation}
\begin{split}\label{eq:sqrd_amps}
\left| \mcM_{SM} \right|^2 & \sim \frac{1}{4}\sin ^2\theta \sin ^2\theta_W  \
+\frac{\mw}{\sqrt{\hat s}}  {\cal F}(\theta, \theta_W) \cos\phi_W \,, \\
\Re\mcM_{SM} \, \mcM_{\varphi q}^{(3)*}& \sim\frac{\hat s}{\Lambda^2} \left[
    \frac{1}{4}\sin ^2\theta \sin ^2\theta_W 
    +\frac{\mw}{\sqrt{\hat s}}  {\cal F}(\theta, \theta_W) \cos\phi_W \right]\,,\\
\Re\mcM_{SM} \, \mcM_{\varphi {\textsc w}}^* &\sim \frac{\sqrt{\hat s}\,\mw}{\Lambda^2} {\cal F}(\theta, \theta_W) \cos \phi_W\,,\\
\Re \mcM_{SM} \, \mcM_{\varphi \widetilde {\textsc w}}^* &\sim \frac{\sqrt{\hat s} \,\mw}{\Lambda^2} {\cal F}(\theta, \theta_W) \sin \phi_W\,,
\end{split}
\end{equation}
where ${\cal F}(\theta, \theta_W) = \left(1-\cos\theta\cos\theta_W\right) \sin\theta  \sin\theta_W$. Since integration over the polar decay angle $\theta_W$ does not destroy the interference terms for $\Ohw$, $\Ohwtil$, we choose for the double differential analysis a binning in the azimuthal angle $\phi_W$ and $p_T^h$, see Section~\ref{sec:binning}.

\begin{figure}[t]
	\centering
	\includegraphics[width=0.5\linewidth]{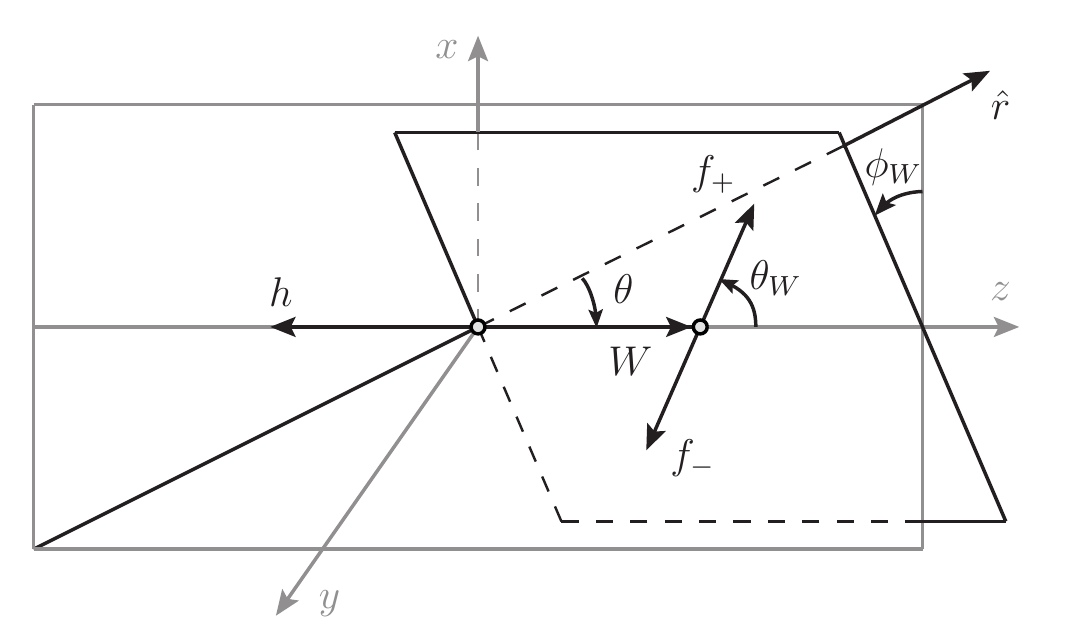}
	\caption{Scattering and decay angles. $f_{\pm}$ denotes the $\pm$ helicity lepton from the $W$ decay.}
	\label{fig:decay_angles}
\end{figure}

The definition of each of the scattering and decay angles is shown in Fig.~\ref{fig:decay_angles}, where the reference vector $\hat r$ is defined as the direction of the boost of the $Wh$ system in the lab frame. 
The scattering angle $\theta$ is defined as the angle between the $W$ boson momentum and $\hat r$ (which for a $2\to2$ process is parallel to the beam axis). 
The positive helicity lepton decay angles, $\theta_W$ and $\phi_W$, are defined in the $W$ rest frame. The positive helicity lepton corresponds to $\ell$ or $\nu$ depending on the sign of the charged lepton. See Ref.~\cite{Panico:2017frx} for more details.

From Eq.~\eqref{eq:sqrd_amps}, we find that the leading interference terms involving $\Ohw$ and $\Ohwtil$ grow with $\sqrt{\hat s}/\Lambda$. This is true provided that we do not integrate over the azimuthal angle, $\phi_W$.
For the $\Ohw$ operator, the leading interference terms after integrating over the azimuthal angle are constant, see Eq.~\eqref{eq:cp-even_interf_nu}.
On the other hand the interference involving $\Ohwtil$ vanishes exactly since the amplitudes have opposite parity.
This is not the case for $\Oqtrip$, since this operator mainly contributes to the
longitudinal $W$ amplitude, which is also the leading SM channel.

There is, however, a subtlety connected to the reconstruction of the neutrino. The missing transverse momentum and the kinematics of the charged lepton can be exploited to reconstruct the neutrino momentum only up to a two-fold ambiguity. This ambiguity, in the limit of high neutrino $p_T$, corresponds to a phase shift,
$\phi_W \rightarrow \pi - \phi_W$~\cite{Panico:2017frx}. 
Because of this, at high energies, all the terms proportional to $\cos \phi_W$ in Eq.~\eqref{eq:sqrd_amps} vanish, while the ones proportional to $\sin \phi_W$ do not. In particular, the leading interference between the $\Ohw$ amplitude and the SM, and the subleading terms in the squared and interference amplitudes for the SM and $\Oqtrip$ average to zero.\footnote{The interference for the $\Ohw$ operator could be restored by considering hadronic $W$ decay channels, in which case the decay angles can be reconstructed up to an ambiguity
$(\theta_W, \phi_W) \rightarrow (\pi - \theta_W, \pi - \phi_W)$~\cite{Panico:2017frx}. This channel, however, has much larger backgrounds, so we do not consider it in our analysis.} On the other hand, the leading interference term for $\Ohwtil$ is unaffected by the ambiguity. It is important to notice that the neutrino ambiguity will not have a significant impact on the bounds of $\Oqtrip$, since as shown in Eq.~\eqref{eq:sqrd_amps} its leading interference with the SM is insensitive to $\phi_W$.
\begin{figure}[t]
	\centering
	\includegraphics[width=0.49\linewidth]{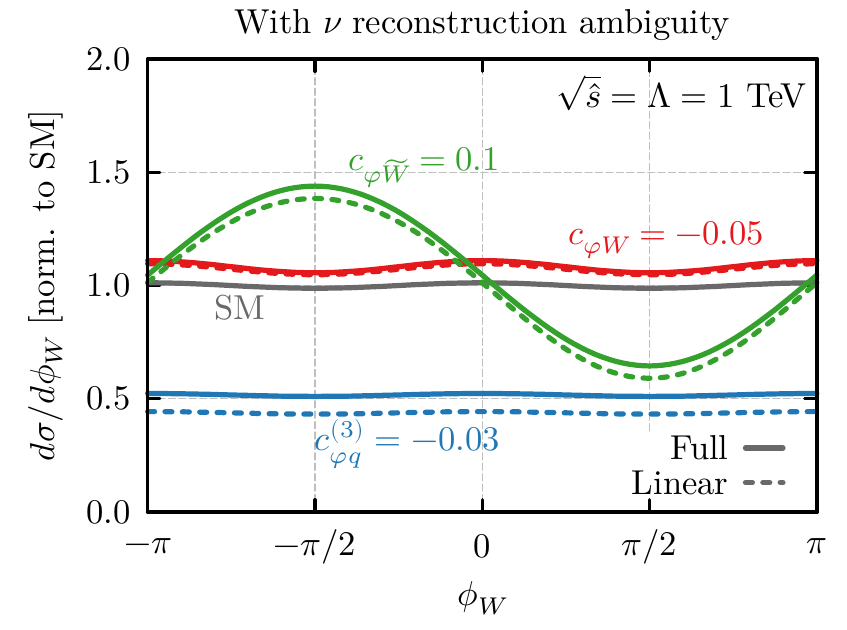}
	\hfill
	\includegraphics[width=0.49\linewidth]{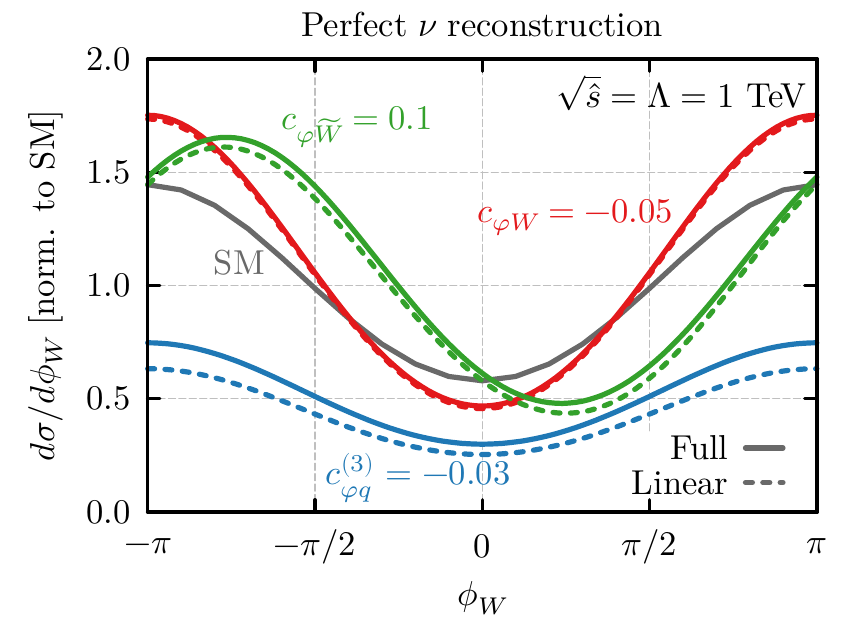}
	\caption{Distributions in the azimuthal angle of the $W$, $\sqrt{\hat s} = 1 \, {\rm TeV}$ and integrated over $\theta$ and $\theta_W$.  Gray lines correspond to the SM distribution, blue to $c_{\varphi q}^{(3)} = - 0.03$, red to $c_{\varphi {\textsc w}} = -0.05$,
	and green to $c_{\varphi \widetilde {\textsc w}} = 0.1$. Dotted lines do not include the squared BSM terms, whereas solid lines correspond to the full amplitude. {\bf Left:} distribution taking into account the ambiguity in the neutrino reconstruction
	($\phi_W \rightarrow \pi - \phi_W$). {\bf Right:} distribution with perfect neutrino reconstruction.
    Note that the values of the Wilson coefficients here were chosen to make the figure legible and are not necessarily within the bounds reported later.	
	}
	\label{fig:distributions}
\end{figure}

After taking into account the neutrino ambiguity, the first non-vanishing interference term for $\Ohw$ does not grow with $\sqrt{\hat s}$ anymore; rather, it is constant. Its explicit analytic expression is:
\begin{equation}\label{eq:cp-even_interf_nu}
\Re\mcM_{SM} \, \mcM_{\varphi {\textsc w}}^* \sim \frac{M_W^2}{\Lambda^2}
\left[(1 - \cos \theta \cos \theta_W)^2 + \frac{1}{2} \sin^2 \theta \sin^2 \theta_W (1 + \cos 2 \phi_W)\right]\,. 
\end{equation}
This expression contains two contributions.
One of them comes from
the interference between the SM and the BSM amplitude with the same $W$ polarization. This term has no $\phi_W$ dependence.
The second contribution comes from an interference between opposite transverse $W$ polarizations. This gives rise to a modulation in the azimuthal angle proportional to $\cos 2 \phi_W$, and vanishes if we integrate over $\phi_W$. A similar modulation in $\cos 2 \phi_W$ can be derived for the SM and $\Oqtrip$ distributions looking at the equations in Appendix \ref{app:amps_squared} and averaging them over the neutrino ambiguity. 

We show the differential distributions with respect to $\phi_W$ for each BSM operator, and the SM, in Fig. \ref{fig:distributions}. We set $\sqrt{\hat s} = 1$ TeV and integrate over all other kinematical variables ($\theta$, $\theta_W$). On the right panel we show the distributions without taking into account the neutrino ambiguity, while on the left panel we plot the same distributions averaging over $\phi_W$ and $\pi - \phi_W$. As discussed above, the SM, $\Ohw$ and $\Oqtrip$ distributions are almost flat when considering the neutrino ambiguity and only show a mild modulation proportional to $\cos 2 \phi_W$, while the $\Ohwtil$ distribution is unchanged and goes like $\sin \phi_W$.

\subsection{Power-counting considerations}\label{sec:power-counting}

Before entering into the actual analysis, we briefly provide some estimates of the size of the Wilson coefficients in common BSM scenarios.

In many models, the largest contributions are expected to be associated to the $\Oqtrip$ operator which can be easily generated at tree-level through the exchange of fermionic partners of the top or vector resonances charged under the EW group. According to the SILH~\cite{Giudice:2007fh} power counting we expect\footnote{Note that in the SILH basis, $c_{\varphi q}^{(3)} = \frac{\Lambda^2}{v^2}(c'_{Hq} + c_W + c_{HW} + 2 c_{2W})$, see Ref.~\cite{Falkowski:2001958}.}
\begin{equation}
c_{\varphi q}^{(3)} \sim g^2\,,
\end{equation}
where $g$ is the EW gauge coupling. This result is valid in theories in which new physics is either weakly coupled or not directly coupled to the SM fields. This happens, for instance, in composite Higgs scenarios, where the new vector resonances interact with the SM fermions through a mixing with the SM gauge fields.

The power counting estimate changes if new physics is strongly coupled to the SM (both to the SM quarks and the Higgs). In this case the estimate becomes
\begin{equation}
c_{\varphi q}^{(3)} \sim g_*^2\,,
\end{equation}
where $g_*$ is the typical size of the new physics coupling. In the fully strongly-coupled case $g_* \sim 4 \pi$.

The $\Ohw$ and $\Ohwtil$ operators are instead typically more suppressed, since they are often generated at loop level~\cite{Giudice:2007fh}.
In the case of weakly coupled new physics one finds
\begin{equation}
c_{\varphi {\textsc w}} \sim c_{\varphi \widetilde {\textsc w}} \sim \frac{g^4}{16 \pi^2}\,.
\label{eq:weakChw}
\end{equation}
If the new physics in the loop are strongly-coupled, then the estimate becomes instead $c_{\varphi {\textsc w}} \sim c_{\varphi \widetilde {\textsc w}} \sim g^2 g_*^2/16\pi^2$. Larger effects in these operators could be present in theories with ``remedios''-like power counting, in which the transverse components of the gauge fields are strongly coupled to new physics~\cite{Franceschini:2017xkh,Liu:2016idz}. In this case the values of the Wilson coefficients could become as large as $c_{\varphi {\textsc w}} \sim c_{\varphi \widetilde {\textsc w}} \sim g_*^4/({16 \pi^2})$.

For completeness, let us mention that $\Ohw$ and $\Ohwtil$ could be generated at tree-level by massive vector fields ${\cal L}_1$ in the ${\bf 2}_{-1/2}$ representation of ${\rm SU}(2)_L \times {\rm U}(1)_Y$, via mixing with the Higgs ${\cal L}_{1\mu}^\dagger D^\mu H + {\rm h.c.}$~\cite{deBlas:2017xtg}.
Moreover, this multiplet does not give rise to a tree-level contribution to $\Oqtrip$. If the ${\cal L}_1$ multiplet is the lightest new physics state, the $\Ohw$ and $\Ohwtil$ operators could therefore receive larger contributions than $\Oqtrip$.
This situation, however, does not arise in the most common BSM scenarios.

In any case where $\Chw$ or $\Chwtil$ is parametrically enhanced with respect to Eq.~\eqref{eq:weakChw}, minimal coupling imposes structural cancellations in the contributions to $h \to \gamma \gamma$~\cite{Giudice:2007fh}. Hence one should be careful when setting bounds on these operators from Higgs data.

\section{Event generation and analysis}\label{sec:signal_background}

Our signal process is $pp\to W(\to\ell\nu)\,h(\to\gamma\gamma$).
The main backgrounds for this process are $W\gamma\gamma$, $W\gamma j$, and $W j j$, with the jets faking a photon.\footnote{We assume that, at the FCC-hh, the rapidity coverage of the detector is large enough such that the background due to $pp\to Z(\to\ell\ell)\,h(\to\gamma\gamma)$ with one missing lepton is negligible.}
We take the rate for a jet to fake a photon to be $P_{j\to\gamma}=10^{-3}$. Moreover, while this rate is conservative with respect to the fake rates reported in~\cite{Contino:2016spe,Abada:2019lih}, see Appendix~\ref{app:mc_evt_gen}, reducing it further does not have a significant impact on the bound we obtain. We simulated the events with \textsc{MadGraph5\_aMC@NLO} v.2.6.5~\cite{Alwall:2014hca} using the \texttt{NNPDF23\_LO} parton distribution functions~\cite{Ball:2013hta}. We used \textsc{Pythia8.2}~\cite{Sjostrand:2014zea} to model the parton shower and to decay the Higgs into two photons. Detector effects were modeled with \textsc{Delphes} v.3.4.1~\cite{deFavereau:2013fsa,Selvaggi:2014mya,Mertens:2015kba,Cacciari:2011ma,Cacciari:2005hq,Cacciari:2008gp} using its FCC-hh card.
For a detailed discussion of the event generation, the applied generation cuts, and QCD and EW radiative corrections, see Appendix~\ref{app:mc_evt_gen}.
\subsection{Selection cuts}
\label{sec:cuts}
To reconstruct the signal, we require at least one electron or muon with $p_T > 30$ GeV, missing
transverse momentum, $\slashed{E}_T>100$ GeV, and at least two photons, each with
$p_T>50$ GeV and with an invariant mass $m_{\gamma \gamma} \in [120,130]$ GeV. If more
than two pairs of photons satisfy this condition, we select the pair with the
smallest distance between the two photons defined as $\Delta R^{\gamma\gamma} =
\sqrt{(\Delta \eta^{\gamma \gamma})^2 + (\Delta \phi^{\gamma \gamma})^2}$.
These selection cuts are summarized in Table~\ref{tab:sel_cuts}.

To further reduce the backgrounds, we apply two additional cuts: first, we impose a maximum separation cut on the diphoton pair, $\Delta R^{\gamma\gamma}_{\max}$;
since the background pair is non-resonant, it tends to have a larger $\Delta R^{\gamma\gamma}$ than the one that originates from the boosted-Higgs.
Second, we impose a maximum cut on the transverse momentum of the reconstructed $Wh$ system,
\begin{equation}
p_T^{Wh} <  p_{T,\max}^{Wh}\,.
\end{equation}
This cut is motivated by the fact that we only expect large $Wh$ transverse momenta from processes recoiling against hard QCD jets and not from the growth with energy we expect from BSM physics.
Moreover, to increase the efficacy of the $\Delta R^{\gamma\gamma}$ and $p_T^{Wh}$  cuts, their values depend on the choice of binning for $\pth$ which is described in the next section.

\begin{table}[t]
	\centering{
		\renewcommand{\arraystretch}{1.25}
		\begin{tabular}{ c @{\hspace{.5em}} | @{\hspace{.5em}} c   }
			\toprule
			& Selection cuts  \\\midrule
			$p_{T,\min}^{\ell}$ [GeV] &  30 \\
			$p_{T, \min}^{\gamma}$  [GeV] & 50 \\
			$\slashed{E}_{T, \, \min}$ [GeV] & 100  \\
			$m_{\gamma \gamma}$ [GeV] & $[120,130]$ \\
			\hline
			$\Delta R^{\gamma\gamma}_{\max}$ & $\{1.3, 0.9, 0.75, 0.6, 0.6\}$\\ 
			$p_{T,\max}^{Wh}$ [GeV] &  $\{300,500,700, 900, 900\}$  \\
			\bottomrule
		\end{tabular}}
		\caption{
		A summary of the selection cuts applied to the Monte Carlo events. The entries for $\Delta R^{\gamma\gamma}_{\max}$ and $p_{T,\max}^{Wh}$ correspond to the cuts performed in each $\pth$ bin, see Section~\ref{sec:binning} for the definition of the bins.
		}
		\label{tab:sel_cuts}
	\end{table}

\subsection{Binning of the double differential distribution}
\label{sec:binning}
%

As discussed in Section~\ref{sec:interference}, to maximize the sensitivity to the three operators of interest we need to select events with large $Wh$ invariant mass and be differential in the azimuthal angle of the leptons, $\phi_W$, in order to have linear sensitivity to the 
CP-odd operator. Since new physics effects are largely in the central scattering region, it is convenient to perform a binning in the $p_T^{h}$ variable, which is also correlated with $\sqrt{\hat s}$.
In our analysis we use the following binning,
\begin{equation}
\begin{split}
p_T^{h} &\in \{200, 400, 600, 800, 1000, \infty\}\,{\rm GeV}\,,\\
\phi_W &\in [-\pi,0]\,,\; [0,\pi]\,.
\end{split}
\label{eq:binning}
\end{equation}
With this choice, the overflow $p_T^h$ bin contains ${\cal O}(10)$ SM events for $30 \, \text{ab}^{-1}$ of integrated luminosity.
The number of events per $p_T^h$ bin after all the cuts is shown in Fig.~\ref{fig.pTh} for both the signal and the backgrounds.

As explained in Section~\ref{sec:interference}, the binning in $\phi_W$ is chosen to enhance the sensitivity to $\Ohwtil$ by recovering the interference with the leading SM amplitude.
On the other hand, no improvement is possible for $\Ohw$ since the neutrino ambiguity
cancels the leading modulation in $\phi_W$, as can be seen in the left panel of Fig.~\ref{fig:distributions}.

To efficiently populate the five $\pth$ bins in Eq.~\eqref{eq:binning}, we generated four runs with a cut on $\pth$ or a proxy for it in the case of backgrounds, see Appendix~\ref{app:mc_evt_gen}.

\subsection{Cut efficiencies}
To evaluate the effectiveness of the selection cuts described in Section~\ref{sec:cuts} in suppressing the backgrounds, we show in Table~\ref{tab.cutflow} the cutflow analysis for the third $\pth$ bin.
Due to small differences in the generation-level cuts (see Table~\ref{tab:gen_cuts}), the starting phase space volumes differ slightly. Nevertheless, the analsyis gives a good idea of which cuts are the most effective in suppressing the backgrounds.
In particular, apart from the efficiency due to the $j\to\gamma$ fake rate, $P_{j\to\gamma}=10^{-3}$, Table~\ref{tab.cutflow} shows the effectiveness of the $m_{\gamma\gamma}$ window cut which reduces the backgrounds by more than one order of magnitude. Furthermore, the $p_T^{Wh}$ cut reduces the $W\gamma\gamma\,(Wj\gamma)$ by a factor of $5\,(3)$ respectively.

\begin{table}
\centering\renewcommand*{\arraystretch}{1.5}
\begin{tabular}{c|>{\centering\arraybackslash}p{0.1\textwidth}|>{\centering\arraybackslash}p{0.1\textwidth}|>{\centering\arraybackslash}p{0.1\textwidth}|>{\centering\arraybackslash}p{0.1\textwidth}}
     \toprule
     \rule[-5pt]{0pt}{.5em}Selection cuts / efficiency  & $\xi_{h \to \gamma \gamma}^{(3)}$ & $\xi_{\gamma \gamma}^{(3)}$ & $\xi_{j \gamma}^{(3)}$ & $\xi_{j j}^{(3)}$ \\
     \hline
     $\geq 1 \ell^\pm$ with $p_{T} > 30$ GeV & $0.86$ & $0.46$ &$0.94$ & $0.94$ \\
     $\geq 2 \gamma$ each with $p_{T}> 50$ GeV& $0.50$ & $0.18$ &$5.7\cdot10^{-3}$ & $8.7\cdot10^{-7}$ \\
     $\slashed{E}_T > 100$\,GeV & $0.49$ & $0.16$ & $5.1\cdot10^{-3}$&  $8.5\cdot10^{-7}$\\
     $120\,\text{GeV} < m_{\gamma\gamma} < 130\,\text{GeV} $ & $0.46$ & $6\cdot10^{-3}$ & $2\cdot10^{-4}$ & $8.2\cdot10^{-8}$ \\
     $\Delta R^{\gamma\gamma}< \Delta R_{max}$ & $0.45$ & $4\cdot10^{-3}$ & $3.1\cdot10^{-5}$ & $6.4\cdot10^{-8}$ \\
     $p_T^{Wh} < p_{T,max}^{Wh}$ & $0.41$ & $7\cdot10^{-4}$ & $1.1\cdot10^{-5}$ & $4.7\cdot10^{-8}$ \\
     \bottomrule
     \end{tabular}%
     \caption{Cut-flow efficiency for the selection cuts. The superscript, $(3)$, refers to the third $\pth$ bin at generation level.
     }
   \label{tab.cutflow}%
\end{table}

\begin{figure}
	\centering
	\includegraphics[width=0.74\linewidth]{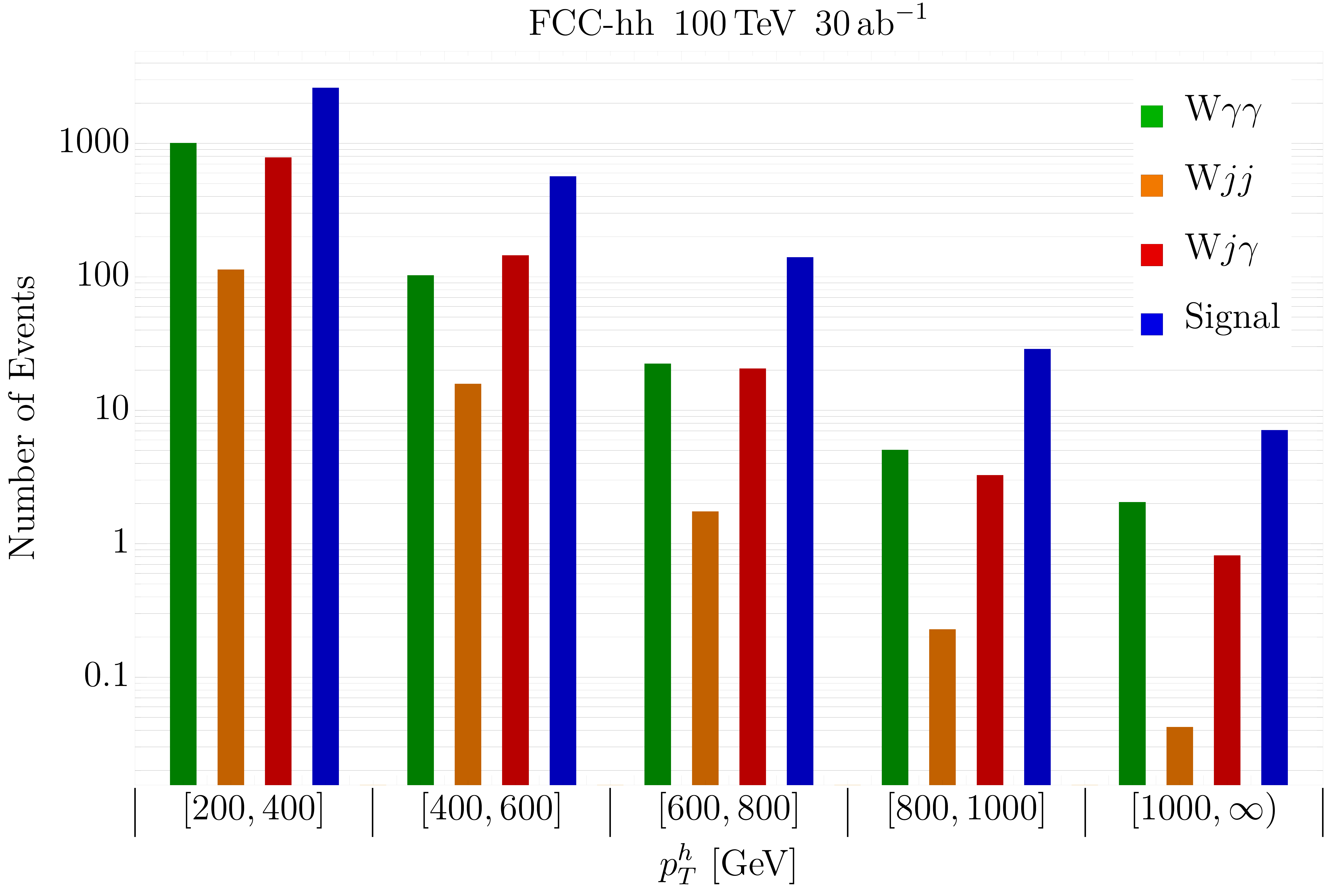}
	\caption{Number of SM events per $p_T^{h}$ bin after selection cuts for the signal and backgrounds at the FCC-hh assuming 30 ab$^{-1}$.}
	\label{fig.pTh}
\end{figure}

The number of SM events in each bin for the signal and background channels after applying the selection cuts are shown in Fig.~\ref{fig.pTh}.
This figure shows that the dominant backgrounds are $W\gamma\gamma$ and $Wj\gamma$. Their size is roughly one third of the signal in the first bin and gradually reduces to roughly $10\%$ of the signal in the last bin.
The $Wjj$ background, however, is at least one to two orders of magnitude smaller than the signal and it can, therefore, be safely neglected.

Let us stress that the exact background projections for $j \gamma$ and $jj$ crucially depend on the jet fake rate into photons and therefore is highly sensitive to the detector performance. Nonetheless, given that even after taking a very conservative estimate for the fake rate the backgrounds are much smaller than the signal; we do not expect our bounds to change much even if the fake rate is further reduced.

In summary, the selection cuts used in our analysis are very efficient in reducing the backgrounds while preserving most of the signal, rendering the $p p \to Wh \to \ell \nu \gamma \gamma$ channel essentially background-free.

\section{Results}
\label{sec.results}
In this section, we present the bounds on the Wilson coefficients of the three operators in question. As a first step,  in Section~\ref{sec:exclusive_fit}, we focus exclusively
on the $\Oqtrip$ operator. This simplification is justified because, in many BSM scenarios, the $\Oqtrip$ operator is generated at tree level while $\Ohw$ and $\Ohwtil$ are generated at loop level, see Section~\ref{sec:power-counting}.
Such an exclusive analysis is also sensible in BSM models where the three Wilson coefficients are of the same size.
In fact, the $\Oqtrip$ operator induces larger deviations in the $Wh$ distributions and can be tested with much higher accuracy than $\Ohw$ and $\Ohwtil$.
We verify this statement quantitatively in Section~\ref{sec:global_fit} where we perform a combined three-operator analysis.

\subsection{Single operator analysis: $\Oqtrip$}\label{sec:exclusive_fit}
As we discussed in Section~\ref{sec:EFT}, the $\Oqtrip$ operator contributes
to the amplitude with a longitudinally-polarized $W$ boson. It directly interferes with the leading SM amplitude giving a quadratic growth with energy. In this case, an analysis strategy with a simple binning in the transverse momentum of the Higgs is adequate to extract the bounds on $c_{\varphi q}^{(3)}$. Moreover, further binning in the leptonic azimuthal angle, $\phi_W$, does not appreciably improve the bound because the leading modulation is destroyed by the neutrino ambiguity.
Nevertheless, to be consistent with the double-differential three-operator analysis that we present in the next subsection, we also use the double binning defined in Eq.~\eqref{eq:binning} for the one-operator analysis.

For the statistical analysis, we assume that the likelihood function is Gaussian for simplicity and do a $\chi^2$ analysis. The Gaussian assumption is justified since we do not have any bins with less than $\mathcal{O}(10)$ events for which the Gaussian approximation is already very good. To construct the $\chi^2$ function, then, we need the number of expected signal events as a function of the Wilson coefficients in each bin. The fits as a function of $\Cqtrip$ are given in Table~\ref{tab.fitperbin} for each $\pth$ bin while the fits in both $\pth$ and $\phi_W$ bins are given in Table~\ref{tab:sigma_full} in Appendix \ref{app:other}.

\begin{table}
\begin{centering}
\begin{tabular}{c|c|c}
\toprule
\rule[-5pt]{0pt}{.5em}
\multirow{2}{*}{$p_T^{h}$ bin} & \multicolumn{2}{c}{Number of expected events} \tabularnewline
\cline{2-3} \cline{3-3} & Signal & Background \tabularnewline
\hline 
\rule[-10pt]{0pt}{2.5em}$[200 - 400]$ GeV & $2620 + 20900\,c_{\varphi q}^{(3)} + 52700\,\big(c_{\varphi q}^{(3)}\big)^{2}$ & $1790$ \tabularnewline
\rule[-10pt]{0pt}{2.5em}$[400 - 600]$ GeV & $566 + 11700\,c_{\varphi q}^{(3)} + 71800\,\big(c_{\varphi q}^{(3)}\big)^{2}$ & $248$ \tabularnewline
\rule[-10pt]{0pt}{2.5em}$[600 - 800]$ GeV & $140 + 5600\,c_{\varphi q}^{(3)} + 67200\,\big(c_{\varphi q}^{(3)}\big)^{2}$ & $43$ \tabularnewline
\rule[-10pt]{0pt}{2.5em}$[800 - 1000]$ GeV & $29 + 1890\,c_{\varphi q}^{(3)} + 36100\,\big(c_{\varphi q}^{(3)}\big)^{2}$ & $8$ \tabularnewline
\rule[-10pt]{0pt}{2.5em}$[1000 - \infty] $ GeV & $7 + 854\,c_{\varphi q}^{(3)} + 33000\,\big(c_{\varphi q}^{(3)}\big)^{2}$ & $3$ \tabularnewline
\bottomrule
\end{tabular}
\par\end{centering}
\caption{Number of expected signal and backgrounds events at FCC-hh with 30 ab$^{-1}$. The signal events number is given as a function of $c_{\varphi q}^{(3)}$ (fixing $\Lambda = 1 \, \text{TeV}$) and the contribution of $Wjj$ to the background is disregarded.
}
\label{tab.fitperbin}
\end{table}

Since we do not know exactly the size of possible systematic errors, we consider three benchmark scenarios with $1\%$, $5\%$, and $10\%$ systematic uncertainty. The $1\%$ benchmark is meant to give an estimate of the maximal sensitivity achievable by removing all possible sources of systematic uncertainty while the $5\%$ one should provide a more realistic assumption, being roughly comparable with the present LHC systematics for production processes with leptonic final states~\cite{Franceschini:2017xkh}. Instead, the $10\%$ benchmark should be considered as a worst-case scenario.

\begin{figure}[t]
	\centering
	\includegraphics[width=0.65\linewidth]{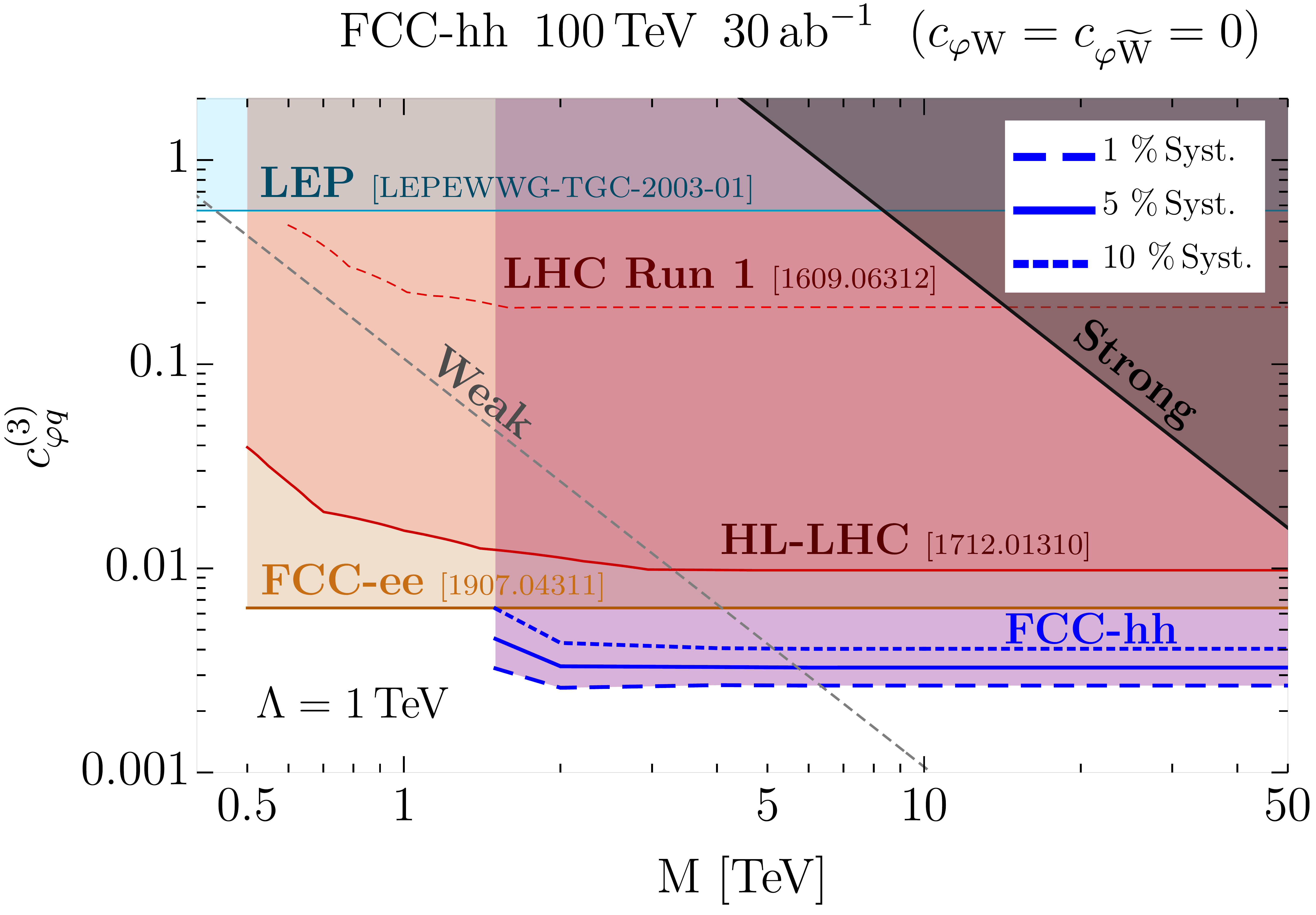}
	\caption{Expected bounds on $c_{\varphi q}^{(3)}$, setting $\Lambda = 1$\,TeV, at the FCC-hh with 30\,ab$^{-1}$ as a function of the maximal invariant mass cut $M$. The bounds correspond to $\Delta \chi^2 = 3.84$ ($\simeq 95\%$ C.L.) and are obtained from a single parameter fit to the $\Oqtrip$ operator. The dashed, solid and dotted blue lines show the bounds for $1\%$, $5\%$ and $10\%$ systematics. The orange shaded area shows the expected bound from a global fit at FCC-ee~\cite{deBlas:2019wgy}, while the shaded red area delimited by a solid (dashed) line corresponds to the HL-LHC with $3 \, \text{ab}^{-1}$  (LHC run 1) bounds from leptonically decaying $WZ$~\cite{Franceschini:2017xkh}. The light blue shaded are corresponds to the bound obtained by LEP~\cite{LEP:2003aa}. The diagonal dashed and solid gray lines show the values of the Wilson coefficient expected in weakly-coupled ($\Cqtrip\sim g^2/(4M^2)$) and strongly-coupled ($\Cqtrip\sim (4\pi)^2/(4M^2)$) new physics models~\cite{Franceschini:2017xkh}.
	}
	\label{fig:fit_cphiq}
\end{figure}

The $\Delta \chi^2 = 3.84$ ($\simeq 95\%$ C.L.) bounds for $\Cqtrip$ are shown in Fig.~\ref{fig:fit_cphiq} for $\Lambda = 1$ TeV, as a function of the cutoff of the EFT, $M$, with 1\%, 5\%, and 10\% systematic uncertainty.
The cutoff is enforced by selecting events with $m_{Wh} \leq M$, where $m_{Wh}$ is the $Wh$ invariant mass, for each value of $M$.\footnote{To reconstruct the invariant mass we randomly chose one of the two neutrino solutions. We checked that this procedure gives the same bound as always choosing for a given event, the solution that gives the highest $\sqrt{\hat s}$, i.e., the conservative case where we reject more events.} This result can give an idea of the
dependence of the bound on the cutoff of the EFT, since we can roughly identify $\Lambda$ with $M$.
The bound saturates for $M \sim 4\,{\rm TeV}$ and there is only a mild degradation for $M \simeq 2\,{\rm TeV}$, below which it rapidly becomes much worse. We find then that for $M \gtrsim 4$ TeV, the $95\%$ C.L.~bounds for $c_{\varphi q}^{(3)}$ (with $\Lambda = 1 \, \text{TeV}$) are:
\begin{equation}\label{eq:oq3_bounds}
\begin{split}
c_{\varphi q}^{(3)} &\in [-2.7, 2.5] \times 10^{-3}\quad 1\%\;{\rm syst.},\\
\rule{0pt}{1.75em}c_{\varphi q}^{(3)} &\in [-3.3, 2.9] \times 10^{-3}\quad  5\%\;{\rm syst.},\\
\rule{0pt}{1.75em}c_{\varphi q}^{(3)} &\in [-4.0, 3.4] \times 10^{-3}\quad  10\%\;{\rm syst.}
\end{split}
\end{equation}
The symmetry between the positive and negative bounds on the Wilson coefficients indicates that either the linear interference term between the SM and BSM dominates the bound or that the squared BSM one does.
One can check that in fact it is the linear term that dominates by comparing the quadratic and linear terms in Table~\ref{tab.fitperbin} setting $c_{\varphi q}^{(3)} \sim {\textit few}\times 10^{-3}\,$.

The diagonal dashed and solid gray lines show the values of the Wilson coefficient that are expected in weakly-coupled new physics models (labeled `Weak' in the plot), with $\Cqtrip\sim g^2/(4M^2)$, and strongly-coupled ones (labeled `Strong'), with $\Cqtrip\sim (4 \pi)^2/(4M^2)$; see discussion in Section~\ref{sec:power-counting}.
The extra factor of $1/4$ is included to match the conventions of Ref.~\cite{Franceschini:2017xkh}; it also arises in the matching of vector-like-quark extensions of the SM, see Ref.~\cite{deBlas:2017xtg}.

For comparison, the projections obtained for the leptonic $WZ$ at the FCC-hh, assuming $5\%$ systematics give a bound (for $M \gtrsim 5$ TeV)~\cite{Franceschini:2017xkh}
\begin{equation}
\label{eq:FCChh_bounds}
\begin{array}{l@{\hspace{.5em}}l@{\hspace{2.em}}l@{\hspace{2.em}}l}
    \textrm{FCC-hh} & {\small (20\;{\rm ab}^{-1})} & c_{\varphi q}^{(3)} \in [-1.8, 1.4] \times 10^{-3}\; & 5\%\;{\rm syst.},
\end{array}
\end{equation}
which is about $2$ times stronger than the one we find from $W(h \to \gamma \gamma)$.\footnote{The $WZ$ channel fit from Ref.~\cite{Franceschini:2017xkh} exploited a simple binned analysis similar to the one used by us. It neglected possible backgrounds which, however, are not expected to be large.}

To give an idea of the FCC-hh constraining power, in Fig.~\ref{fig:fit_cphiq} we also show the current bounds from LEP and LHC run 1 and the expected bounds from leptonic $WZ$ production at HL-LHC~ and from a global fit at future lepton colliders. We summarize these bounds in the following:
\begin{equation}\label{eq:bounds_1}
\def\arraystretch{1.3}
\begin{array}{l@{\hspace{.8em}}l@{\hspace{2.em}}l@{\hspace{2.em}}l}
\textrm{LEP \cite{LEP:2003aa}} &  & c_{\varphi q}^{(3)} \in [-5.7, 5.7] \times 10^{-1}\;, \\
\textrm{HL-LHC \cite{Franceschini:2017xkh}} & {\small (3\;{\rm ab}^{-1})} & c_{\varphi q}^{(3)} \in [-1.2, 1.0] \times 10^{-2}\; & 5\%\;{\rm syst.}, \\
\textrm{HE-LHC \cite{Franceschini:2017xkh}} & {\small (27\;{\rm TeV},\, 10\;{\rm ab}^{-1})} & c_{\varphi q}^{(3)} \in [-4.0, 3.3] \times 10^{-3}\; & 5\%\;{\rm syst.},\\
\textrm{ILC \cite{deBlas:2019wgy}} & {\small} & c_{\varphi q}^{(3)} \in [-6.0, 6.0] \times 10^{-3}\;, \\
\textrm{CEPC \cite{deBlas:2019wgy}} & {\small} & c_{\varphi q}^{(3)} \in [-1.0, 1.0] \times 10^{-2}\;, \\
\textrm{CLIC \cite{deBlas:2019wgy}} & {\small} & c_{\varphi q}^{(3)} \in [-8.0, 8.0] \times 10^{-3}\;, \\
\textrm{FCC-ee \cite{deBlas:2019wgy}} & {\small} & c_{\varphi q}^{(3)} \in [-6.4, 6.4] \times 10^{-3}\;, \\
\end{array}
\end{equation}
where the lepton collider bounds cited above are valid for $M \gtrsim M_Z$ while the hadron collider ones only for $M \gtrsim 5$ TeV \cite{Franceschini:2017xkh}.

Comparing these results with the ones in Eqs.~(\ref{eq:oq3_bounds}) and~\eqref{eq:FCChh_bounds}, we find that the HE-LHC projections are slightly weaker than ours, and the FCC-ee ones are worse by a factor $\sim 3$. It must, however, be noted that the FCC-ee center of mass energy is much lower than the one at FCC-hh and other hadron colliders. Therefore, the corresponding bound on $c_{\varphi q}^{(3)}$ is also valid for low cutoffs, which cannot be tested at hadron machines.

\begin{figure}[t]
	\centering
	\includegraphics[width=0.47\linewidth]{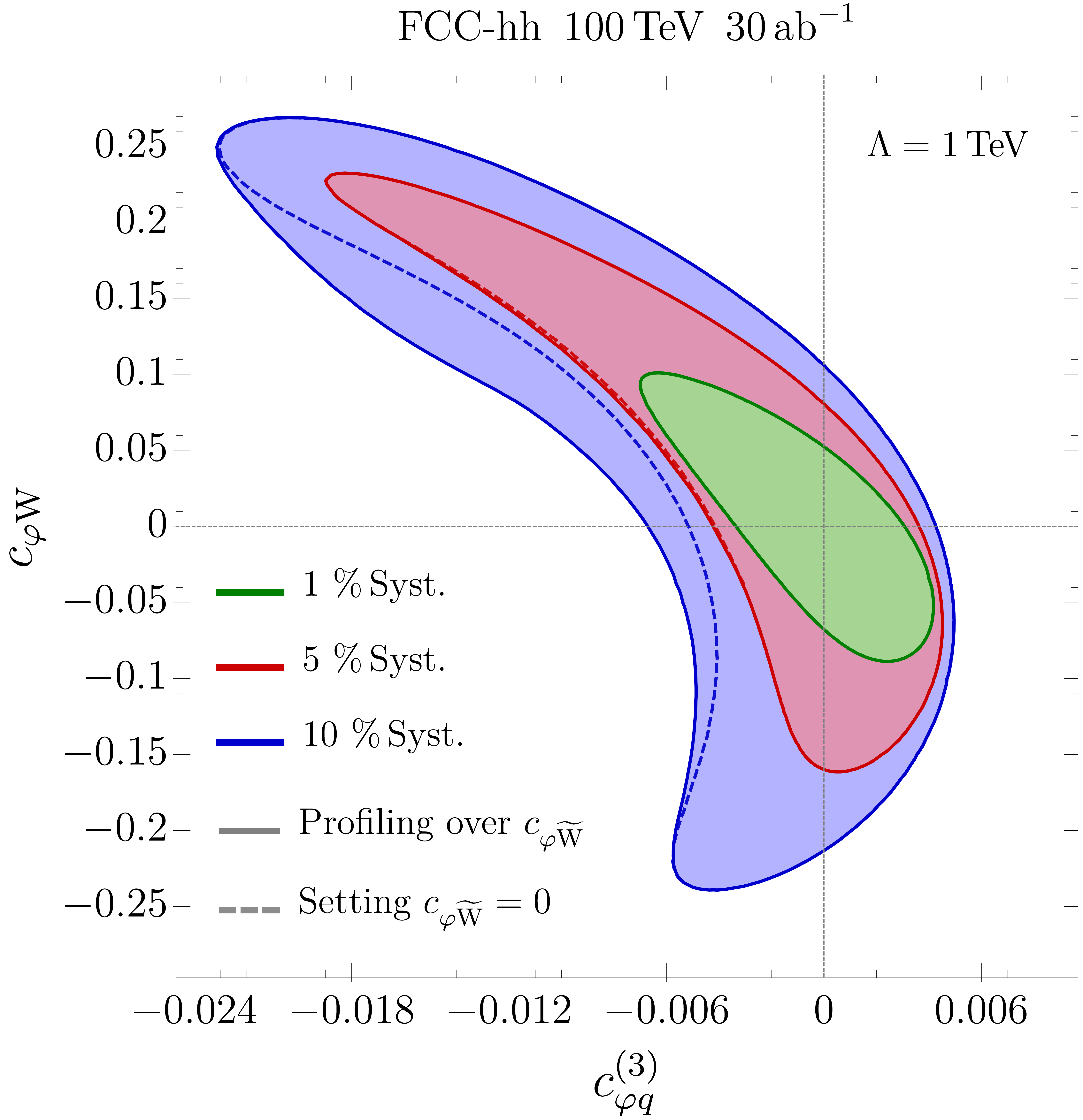} 
	\includegraphics[width=0.47\linewidth]{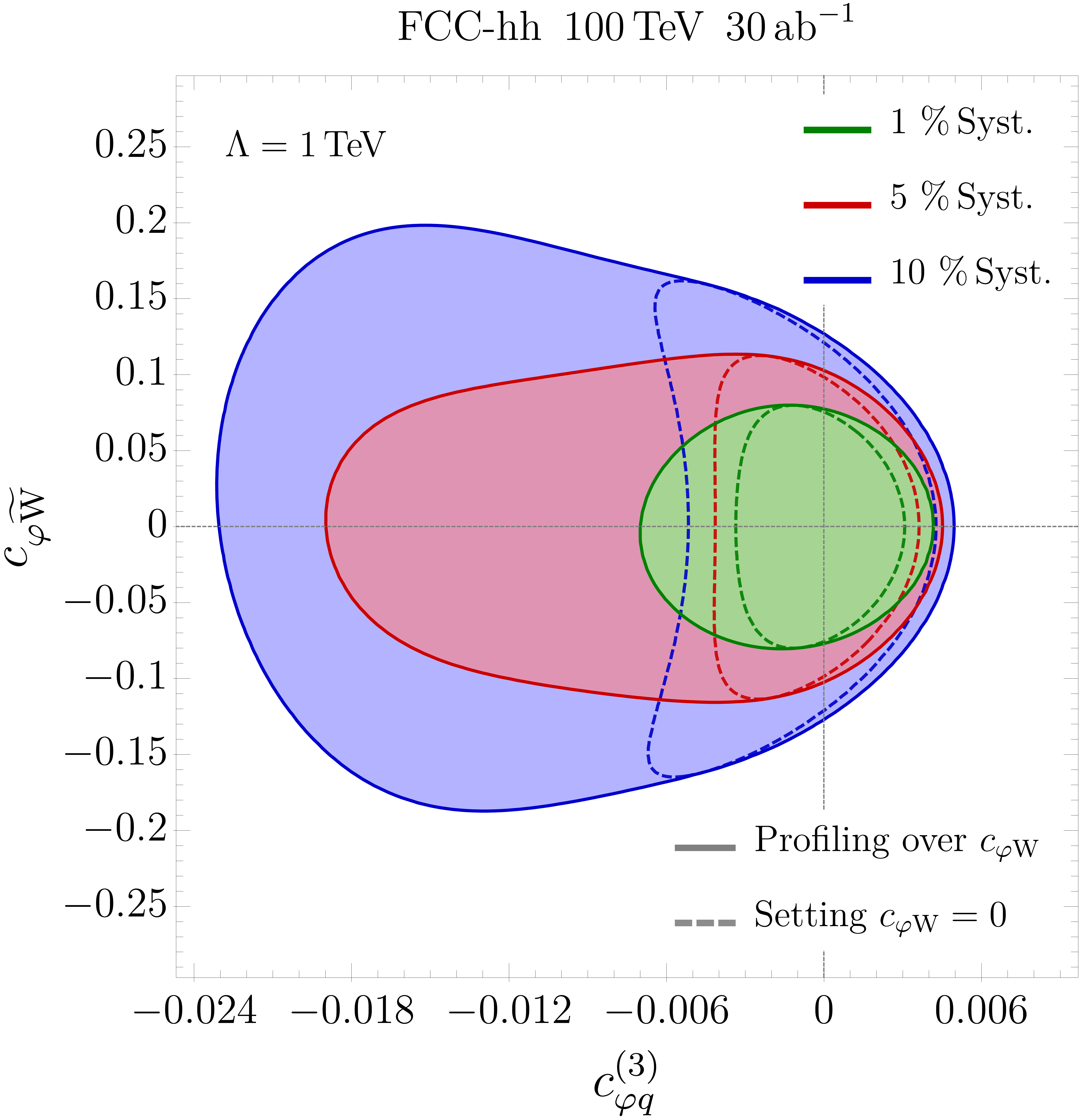} \\
	\vspace{0.3cm}
    \includegraphics[width=0.47\linewidth]{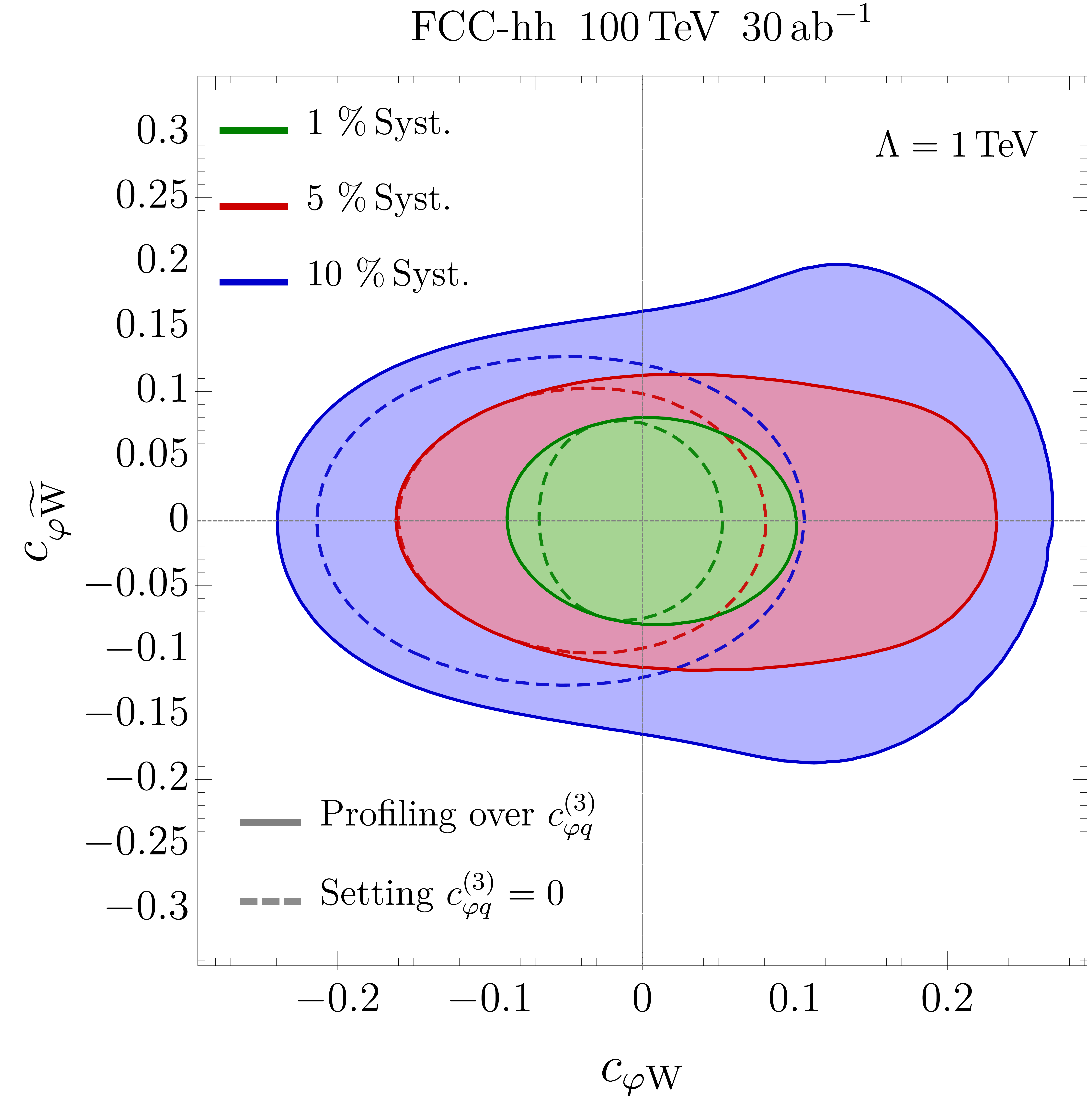}
	\includegraphics[width=0.47\linewidth]{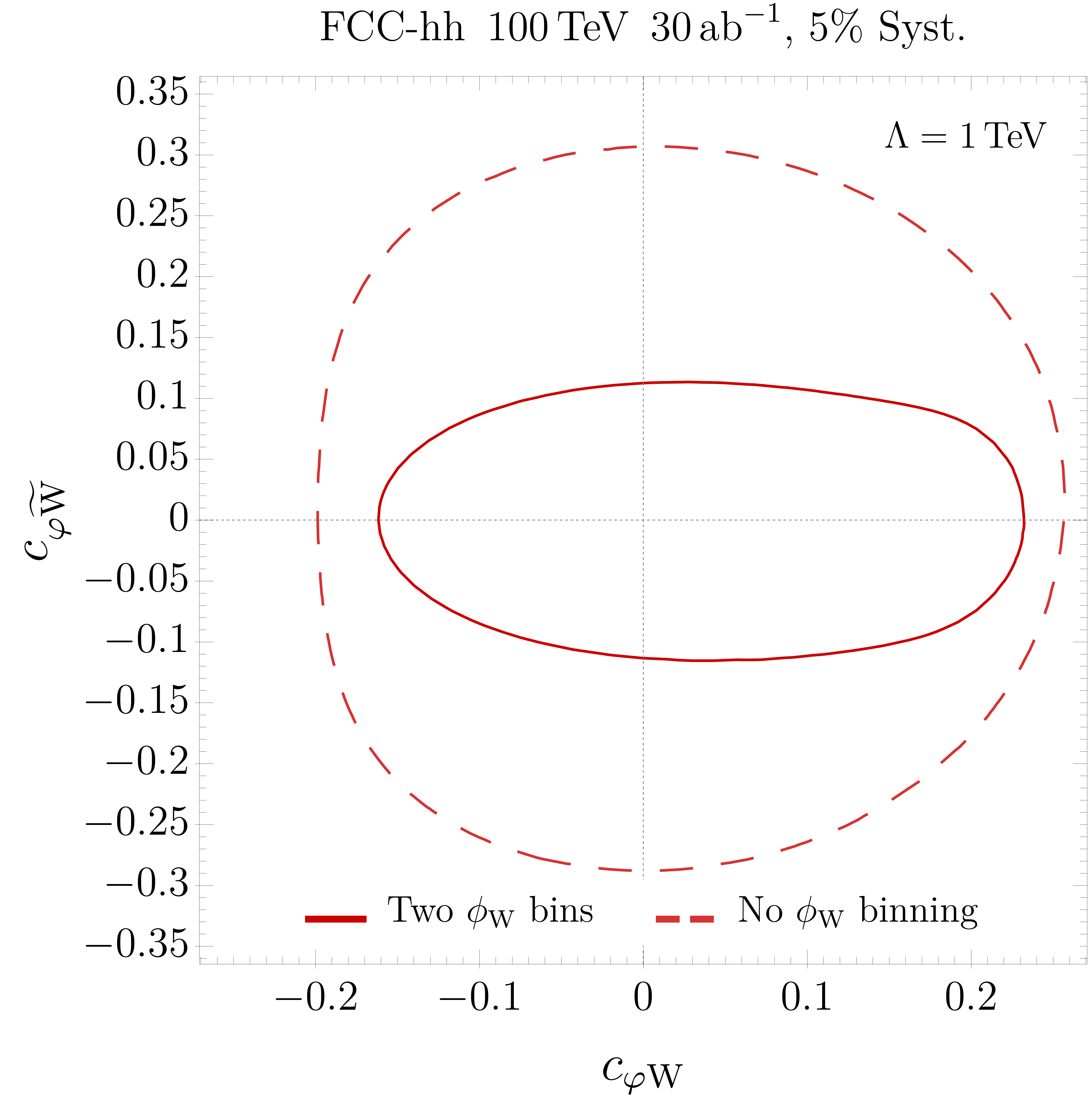}
	\caption{Expected $95\%$ C.L. bounds on $\Chw$, $\Chwtil$, $\Cqtrip$ at the FCC-hh for $30 \, \text{ab}^{-1}$. Bounds in green, red, blue, assume 1\%, 5\% and 10\% systematic error. Solid (dashed) lines in top and bottom left panels correspond to the bounds when profiling over (setting to zero) the Wilson coefficient not appearing in the plot {\bf Top Left:} Bounds on the $\Chw$, $\Cqtrip$ plane. {\bf Top Right:} Bounds on the $\Chwtil$, $\Cqtrip$ plane.  {\bf Bottom Left:} Bounds on the $\Chw$, $\Chwtil$ plane. {\bf Bottom Right:} In solid red (dashed red) the bounds when using (not using) the double differential binning in $\phi_W$ assuming $5\%$ systematics in the plane $\Chw, \, \Chwtil$.
	}
	\label{fig:bounds2d_cphi}
\end{figure}

\subsection{Three operator analysis: {\normalfont $\Ohw$}, {\normalfont $\Ohwtil$}, $\Oqtrip$}\label{sec:global_fit}
We now extend our analysis to all the three operators, $\Oqtrip$, $\Ohw$, and $\Ohwtil$, simultaneously.
The number of events in each bin as a function of the Wilson coefficients is reported in Table~\ref{tab:sigma_full} in Appendix~\ref{app:other}. 

The $95\%$ C.L. constraints
in the $(c_{\varphi q}^{(3)} - c_{\varphi {\textsc w}})$,
$(c_{\varphi q}^{(3)} - c_{\varphi \widetilde {\textsc w}})$, and $(c_{\varphi {\textsc w}} - c_{\varphi \widetilde {\textsc w}})$ planes are shown in Fig.~\ref{fig:bounds2d_cphi} in the top-left, top-right, and bottom-left panels respectively. The bottom-right panel of the figure shows the effect of binning in $\phi_W$ on the bound on $\Chwtil$.
We present two sets of results here. The first one is obtained by profiling over the additional Wilson coefficient and is delineated by solid contours in the plots. The second set of results, delineated by dashed contours, is obtained by setting the remaining Wilson coefficient to zero. All the results are given for the three benchmark scenarios with $1\%$, $5\%$ and $10\%$ systematic uncertainty (green, red, and blue contours, respectively).

As expected, the constraints on $c_{\varphi q}^{(3)}$ are stronger, by roughly one order of magnitude, than the ones on $c_{\varphi {\textsc w}}$ and $c_{\varphi \widetilde {\textsc w}}$. This confirms the
expectation that a one-operator analysis for $c_{\varphi q}^{(3)}$ is fully justified even in BSM scenarios in which 
the contributions to all three effective operators are of the same order.

The top left panel in Fig.~\ref{fig:bounds2d_cphi} shows that the $c_{\varphi q}^{(3)}$ and $c_{\varphi {\textsc w}}$ operators are correlated, mainly due to the fact that both operators can only be distinguished by a different growth in the $p_T^{h}$ distribution.
On the other hand, $c_{\varphi \widetilde {\textsc w}}$ is basically uncorrelated with the other two Wilson coefficients.
This is because the linear sensitivity to $c_{\varphi \widetilde {\textsc w}}$ derives from the $\phi_W$ binning which allows for interference with the SM.
The effect of the $\phi_W$ binning can be clearly seen in the bottom right panel of Fig.~\ref{fig:bounds2d_cphi}.
For the 5\% systematic uncertainty benchmark, the bound is $\sim 3$ times stronger with only two $\phi_W$ bins in comparison with no binning. The improvement is even larger for the 1\% systematic uncertainty case.
Meanwhile, the binning in $\phi_W$ has a very mild effect on the bound on $c_{\varphi {\textsc w}}$.

The impact of the systematic error on the fits is also quite strong. For $1\%$ systematics, the bounds are mainly driven by the linear SM-BSM interference terms in the cross section for all three operators.
However, for the $5\%$ and $10\%$ benchmarks, quadratic terms clearly play an important role, significantly worsening and distorting the constraints.

It is also interesting to compare the results obtained by our two analysis procedures, i.e. profiling over versus setting to zero the remaining Wilson coefficient. From the upper left plot in Fig.~\ref{fig:bounds2d_cphi},
one can see that the constraints in the $(c_{\varphi q}^{(3)} - c_{\varphi {\textsc w}})$
plane remain almost unchanged. This is expected, since the correlation between these Wilson coefficients and $c_{\varphi \widetilde {\textsc w}}$
is very small. On the contrary, the correlation between $c_{\varphi q}^{(3)}$ and $c_{\varphi {\textsc w}}$ leads to a significant weakening of the bounds on each of these coefficients when we profile over the other one. 
Profiling over $c_{\varphi q}^{(3)}$ or $c_{\varphi {\textsc w}}$ has instead only a minor impact on the determination of $c_{\varphi \widetilde {\textsc w}}$.

\begin{table}
\begin{centering}
\begin{tabular}{c|c|c}
\toprule
Coefficient & Profiled Fit & One Operator Fit \tabularnewline
\hline 
$c_{\varphi q}^{(3)}$ &
\begin{tabular}{ll}
\rule{0pt}{1.25em}$[-5.1,\,3.4]\times10^{-3}$ & $1\%$ syst.\\
\rule{0pt}{1.25em}$[-11.6,\,3.8]\times10^{-3}$ & $5\%$ syst.\\
\rule[-.65em]{0pt}{1.9em}$[-20.6,\,4.1]\times10^{-3}$ & $10\%$ syst.
\end{tabular}
&
\begin{tabular}{ll}
\rule{0pt}{1.25em}$[-2.7,\,2.5]\times10^{-3}$ & $1\%$ syst.\\
\rule{0pt}{1.25em}$[-3.3,\,2.9]\times10^{-3}$ & $5\%$ syst.\\
\rule[-.65em]{0pt}{1.9em}$[-4.0,\,3.5]\times10^{-3}$ & $10\%$ syst.
\end{tabular}
\tabularnewline

\hline
$c_{\varphi {\textsc w}}$ &
\begin{tabular}{ll}
\rule{0pt}{1.25em}$[-7.1,\,7.9]\times10^{-2}$ & $1\%$ syst.\\
\rule{0pt}{1.25em}$[-13.0,\,17.5]\times10^{-2}$ & $5\%$ syst.\\
\rule[-.65em]{0pt}{1.9em}$[-20.0,\,25.2]\times10^{-2}$ & $10\%$ syst.
\end{tabular}
&
\begin{tabular}{ll}
\rule{0pt}{1.25em}$[-5.3,\,4.3]\times10^{-2}$ & $1\%$ syst.\\
\rule{0pt}{1.25em}$[-12.1,\,6.8]\times10^{-2}$ & $5\%$ syst.\\
\rule[-.65em]{0pt}{1.9em}$[-18.8,\,9.0]\times10^{-2}$ & $10\%$ syst.
\end{tabular}
\tabularnewline

\hline
$c_{\varphi \widetilde {\textsc w}}$ &
\begin{tabular}{ll}
\rule{0pt}{1.25em}$[-6.4,\,6.4]\times10^{-2}$ & $1\%$ syst.\\
\rule{0pt}{1.25em}$[-9.0,\,8.8]\times10^{-2}$ & $5\%$ syst.\\
\rule[-.65em]{0pt}{1.9em}$[-13.5,\,14.2]\times10^{-2}$ & $10\%$ syst.
\end{tabular}
&
\begin{tabular}{ll}
\rule{0pt}{1.25em}$[-6.1,\,6.1]\times10^{-2}$ & $1\%$ syst.\\
\rule{0pt}{1.25em}$[-8.1,\,8.1]\times10^{-2}$ & $5\%$ syst.\\
\rule[-.65em]{0pt}{1.9em}$[-10.1,\,10.1]\times10^{-2}$ & $10\%$ syst.\\
\end{tabular} \\
\bottomrule

\end{tabular}
\par\end{centering}
\caption[caption]{Bounds at $95\%$ C.L.~on the coefficients of the $\Oqtrip$, $\Ohw$ and $\Ohwtil$ operators setting $\Lambda = 1 \, \text{TeV}$. {\bf Left column:} bounds profiling over the other two coefficients. {\bf Right column:} bounds with a one operator fit, i.e. setting the other two coefficients to zero.}
\label{tab:bounds_summary}
\end{table}

To get a quantitative idea of the sensitivity to each Wilson coefficient, we report in
Table~\ref{tab:bounds_summary} the bounds on $c_{\varphi q}^{(3)}$, $c_{\varphi {\textsc w}}$ and
$c_{\varphi \widetilde {\textsc w}}$ at the FCC-hh with 30 ab$^{-1}$. We list both the fits obtained through profiling and the ones that take into account each operator separately.

We already compared the one-operator fit bounds on $c_{\varphi q}^{(3)}$ with the ones from other colliders in the previous subsection. Here we notice that the bounds from the profiled fit become significantly worse especially for negative values of the Wilson coefficient, whereas they are relatively stable for positive values. For the $5\%$ systematics benchmark, the profiled bounds are worse than the
ones expected from the $WZ$ channel at HE-LHC and roughly comparable with the FCC-ee ones, see Eq.~(\ref{eq:bounds_1}).

For completeness, we compare our constraints on $\Chw$ and $\Chwtil$ with projected and current bounds. We compare our constraints on $\Chw$ with the projections at the HL-LHC, future lepton colliders and FCC-hh. At 95\% C.L., the bounds are expected to be~\cite{deBlas:2019wgy,deBlas:2019rxi}
\begin{equation}
\begin{array}{l@{\hspace{.8em}}l@{\hspace{3.em}}l@{\hspace{2.em}}l}
\textrm{HL-LHC } \ &{\small (3\;{\rm ab}^{-1})}&c_{\varphi \textsc{w}} \in [-0.4, 0.4]\,,  \\
\textrm{FCC-ee / CEPC / ILC}& &c_{\varphi \textsc{w}}  \in [-0.02, 0.02]\,,\\
\textrm{CLIC}& &c_{\varphi \textsc{w}}  \in [-0.01, 0.01]\,,\\
\textrm{FCC-hh}&{\small (30\;{\rm ab}^{-1})} &c_{\varphi \textsc{w}}  \in [-0.01, 0.01]\,,
\end{array}
\end{equation}
for $\Lambda = 1$ TeV, which for future lepton colliders  and FCC-hh are significantly stronger than our results.\footnote{Recall that in minimally coupled models, large single-operator contributions to $h \to \gamma \gamma$ are structurally correlated, cancelling their contribution.}

The situation is very different for $\Ohwtil$, which can be indirectly tested
through the contributions it induces to the electric dipole moment of the electron. Barring accidental cancellations with the contributions from other CP-violating operators, the current experimental results give a constraint $c_{\varphi \widetilde{\textsc w}} \lesssim 2 \cdot 10^{-5}$(with $\Lambda = 1$ TeV)~\cite{Panico:2018hal}, which is three orders of magnitudes stronger than our bound.
The current direct bounds on $\Ohwtil$ at the LHC (with $36$ fb$^{-1}$) are $|c_{\varphi \widetilde{\textsc w}}| \lesssim 11$
and are expected to reach the level $|c_{\varphi \widetilde{\textsc w}}| \lesssim 1$ (with $\Lambda = 1$ TeV) at the HL-LHC~\cite{Bernlochner:2018opw,biektter2020constraining}, which is worse than the bound we obtain. Nonetheless, we expect that the extrapolation of this differential analysis to FCC-hh will overpass the bound derived from our $WH$ analysis.

\subsection{Connection to aTGCs}
\label{sec:Higgs_EDM_aTGC}

We mention that $\Cqtrip$ can be written as a combination of vertex corrections and the anomalous triple gauge coupling, $\delta g_{1z}$
\begin{equation}
    \Cqtrip = \frac{\Lambda^2}{m_W^2} g^2(\delta g_L^{Zu} - \delta g_L^{Zd} - c_\theta^2 \, \delta g_{1z}) \,,
\end{equation}
where $c_{\theta}$ is the cosine of the Weinberg angle. Therefore, for theories where the vertex corrections are small, the bound on $\Cqtrip$ can be recast as a bound on  $\delta g_{1z}$. For universal theories, where $\delta g_L^{Zu}$, $\delta g_L^{Zd}$ depend only on a combination of the oblique parameters $S$, $T$, $W$, $Y$~\cite{Franceschini:2017xkh, Grojean:2018dqj}, this is especially justified, since the oblique parameters are expected to be constrained with excellent accuracy through a variety of measurements at the FCC-ee and FCC-hh, making $\delta g_L^{Zu}$, $\delta g_L^{Zd}$ negligible.

In this way, from the one-operator fit assuming 5\% systematics and setting $\Lambda = 1$~TeV, we obtain
\begin{equation}\label{eq:TGCbound_exclusive}
\delta g_{1z} \in [-5.7, 6.5] \times 10^{-5}\,,
\end{equation}
whereas from the profiled bound we get
\begin{equation}\label{eq:TGCbound_profiled}
\delta g_{1z} \in [-7.5, 22.9]\times 10^{-5}\,.
\end{equation}
For comparison, we collect the current and future estimated bounds on $\delta g_{1z}$:
\begin{equation}
\begin{array}{l@{\hspace{.8em}}l@{\hspace{3.em}}l@{\hspace{2.em}}l}
\textrm{LEP \cite{ALEPH:2003tss}} \ & & \delta g_{1z} \in [-5.1, 3.4]\times 10^{-2}\,,  \\
\textrm{LHC \cite{Grojean:2018dqj}}& &\delta g_{1z}  \in [-15, 1]\times 10^{-3}\,,\\
\textrm{HL-LHC \cite{Franceschini:2017xkh}}& &\delta g_{1z}  \in [-1, 1]\times 10^{-3}\,,\\
\textrm{FCC-ee \cite{deBlas:2019wgy}}& &\delta g_{1z}  \in [-5, 5]\times 10^{-4}\,,
\end{array}
\end{equation}
for $\Lambda = 1$ TeV. The bounds from LHC and HL-LHC were obtained using the diboson production process $pp\rightarrow WV$, with $V=W,\,Z$. It is clear that our results improve the existent and HL-LHC bounds and are even better than the expected bound from FCC-ee.
\section{Summary and conclusions} \label{sec:conclusions}
%
\begin{figure}[t]
	\centering
	\includegraphics[width=0.8\linewidth]{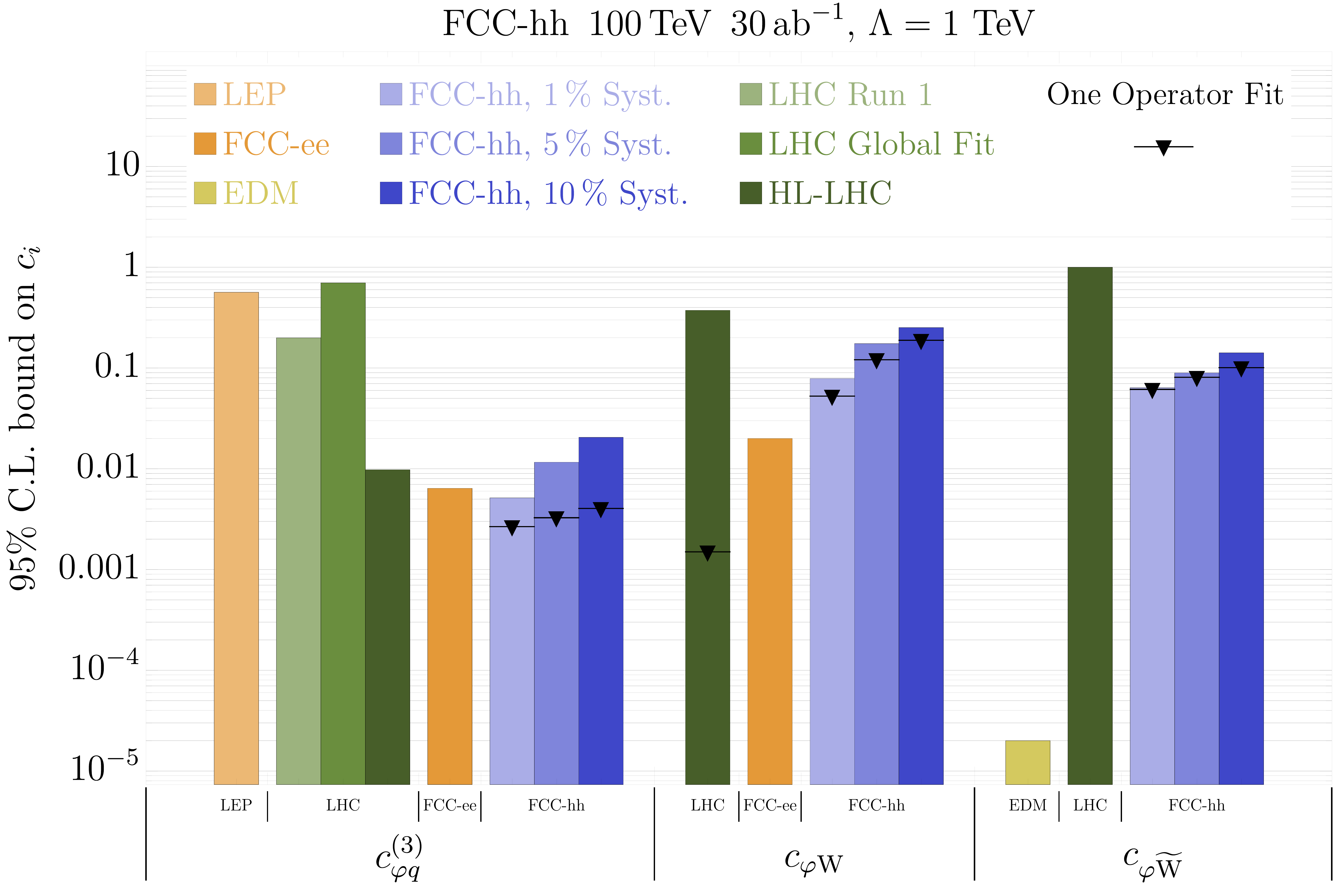}
	\caption{$95\%$ C.L bounds on $c_{\varphi q}^{(3)}$, $c_{\varphi {\textsc w}}$ and $c_{\varphi \widetilde {\textsc w}}$. In blue our bounds from $Wh \to \ell \nu \gamma \gamma$ for FCC-hh with 30 ab$^{-1}$ for different systematics for a three operator fit. The black lines with a triangle on top represent the the bound for a one operator fit instead. In light orange the current LEP~\cite{Franceschini:2017xkh} bound for $c_{\varphi q}^{(3)}$. In lighter and darker green for $\Cqtrip$, the LHC and HL-LHC bounds using leptonic $WZ$ (5\% syst)~\cite{Franceschini:2017xkh}. In medium green the current bounds on $\Cqtrip$ from a global fit~\cite{Ellis:2018gqa}. In darker green also, the HL-LHC projections for $\Chw$~\cite{deBlas:2019wgy,deBlas:2019rxi} (obtained from a global fit) and $\Chwtil$~\cite{Bernlochner:2018opw,biektter2020constraining} (obtained from CP-odd observables in single Higgs production). Assuming no structural cancellations, we show the bound on $\Chw$ from a one parameter fit using $\delta \kappa_\gamma$ in~\cite{deBlas:2019rxi,Cepeda:2019klc} as a triangle on the green bar. In darker orange the FCC-ee bounds from a global fit in~\cite{deBlas:2019wgy} with the configuration FCC-ee Z/WW/$240$~GeV/$365$~GeV. In yellow the EDM bound taken from~\cite{Panico:2018hal}.
 }
	\label{fig:comparison_bounds}
\end{figure}
%

In this work, we analyzed the $p p \to Wh \to \ell \nu \gamma \gamma$ channel at FCC-hh as a way
to perform precision EW measurements. We focused on new physics effects that grow with energy,
parametrized by dimension-6 effective operators within the SMEFT framework. In particular, we
identified three operators that induce a growth with energy in the amplitude. Adopting the
Warsaw basis convention, they are: $\Oqtrip$, which induces a $\hat s$ growth in the longitudinally polarized amplitude, and
$\Ohw$ and $\Ohwtil$, which induce a $\sqrt{\hat s}$ growth in the transverse ones.

We found that a simple analysis strategy exploiting a binning in the $p_T$ of the Higgs boson
is already sufficient to obtain a good sensitivity to $\Oqtrip$.
Testing the $\Ohw$ and $\Ohwtil$ operators is more challenging, since they do not interfere with the
leading SM amplitude if we simply use a $p_T^h$ binning.
We found, nonetheless, that a double
differential distribution which also takes into account the azimuthal decay angle of the lepton, $\phi_W$, can be used to recover the interference for $\Ohwtil$ and significantly strengthen the
bounds. The double binning, however, only has a minor impact on the determination of $\Ohw$.

We show a summary of the projected 95\% C.L. bounds on the three operators in Fig.~\ref{fig:comparison_bounds}. We give two sets of results. The first, given by the blue
bars, corresponds to the constraints derived from the profiling of a fit including all three effective operators. The second set, shown by the black lines with a triangle on top, corresponds
to the fits including one operator at a time. For each of our fits we consider three benchmark scenarios characterized by different systematic uncertainties: $1\%$ (lighter shading), $5\%$ (medium shading) and $10\%$ (darker shading).
One can see that systematics play a significant role in the bounds obtained. We believe that the $5\%$
benchmark could be a realistic estimate, since it is comparable to the present LHC systematics for similar processes with leptonic final states.

Another interesting feature of our results is the impact of a three-operator fit instead of a single-operator one. We find that the bounds on $\Ohw$ and $\Ohwtil$ are almost unchanged, whereas the bound on $\Oqtrip$ is strongly affected. This feature comes from a correlation between $c_{\varphi q}^{(3)}$ and $c_{\varphi \textsc{w}}$. However, it is important to stress that, in a majority of BSM scenarios, the deformations parametrised by $\Ohw$ and $\Ohwtil$ are expected to be subleading with respect to the
ones to $\Oqtrip$. In this class of theories, the correct bound on $\Oqtrip$ is the one obtained from the one operator fit.

Regarding the comparison with present and future bounds from other collider experiments, we find that
the one-operator $\Oqtrip$ constraints are significantly stronger than the ones achievable at the HL-LHC through leptonic $WZ$ production. They are also marginally better than the expected ones at HE-LHC and FCC-ee. The $WZ$ production channel at FCC-hh could instead provide stronger bounds,
although only by a factor~$\sim 2$, see~Eq.~\eqref{eq:FCChh_bounds}.

We find that our bounds on $\Ohw$ are competitive with a projected global fit to Higgs data anticipated by the end of the HL-LHC but not at FCC-ee nor FCC-hh.

Our results for the $\Ohwtil$ determination are one order of magnitude stronger than the ones
achievable at the HL-LHC obtained using CP-sensitive observables from VBF and gluon fusion \cite{Bernlochner:2018opw, biektter2020constraining}. One feature of these observables is that they are linearly sensitive to CP-odd operators, as opposed to inclusive Higgs data. This is also true for the observable we constructed in our analysis. Nevertheless, indirect constraints coming from the current electron EDM bounds are already three orders of magnitude stronger than the expected bound we find.

We conclude by mentioning a few connected research directions that are worth exploring as a continuation of this work. At present we focused on the rare final state $h \rightarrow \gamma \gamma$ since it
provides a cleaner and simpler to analyze channel. Other final states with larger cross section, chiefly among them $h \rightarrow b\bar{b}$, are, however, worth investigating. These channels are more challenging due to the much larger backgrounds, however they could provide access to a significantly larger energy range, allowing us to exploit more efficiently the energy-growing new physics effects.

It is also worth mentioning that $Wh$ is not the only channel that can be exploited for precision measurements. A closely related one is $Zh$ production, that has similar features at FCC-hh. Although this channel has a smaller cross section and is expected to provide weaker bounds on $\Oqtrip$, see Ref.~\cite{Banerjee:2018bio}, it allows us to access an additional operator that induces an $\hat s$ growth,
namely ${\cal O}_{\varphi q}^{(1)}$~\cite{Franceschini:2017xkh}.

\section*{Acknowledgments}

We thank Jorge de Blas, Jiayin Gu, Rick S. Gupta, Ayan Paul, J\"urgen Reuter, and Raoul R{\"o}ntsch for useful discussions. M.M.~was supported by the Swiss National Science Foundation, under Project Nos.~PP00P2176884. G.P.~was supported in part by the MIUR under contract 2017FMJFMW (PRIN2017). F.B.,~P.E.,~C.G.~and A.R.~acknowledge support by the Deutsche
Forschungsgemeinschaft (DFG, German Research Foundation) under Germany’s Excellence
Strategy – EXC 2121 “Quantum Universe” – 390833306. F.B. was also supported by the ERC Starting Grant NewAve (638528). This work was partially performed at the Aspen Center for Physics, which is supported by National Science Foundation grant PHY-1607611.


\appendix

\section{Helicity amplitudes}
\label{app:HelAmps}
In this appendix we write down the explicit formulas for the SM and BSM helicity amplitudes. 

\subsection{$p p \to Wh$}
\label{app:amps_wh}

The $pp\to Wh$ helicity amplitudes given in this subsection are exact. For convenience, we define $\varepsilon_W\equiv \mw/\rtsha$ and $\varepsilon_H\equiv\mh/\rtsha$. The scattering angle $\theta$ is defined in Section~\ref{sec:interference}, see Fig.~\ref{fig:decay_angles}. The $W$ boson polarization vectors are defined with respect to the null reference momentum $(|\vec{p}_W|,\,-\vec{p}_W)$, where $p_W$ is the $W$ momentum in the $Wh$ center of mass frame.
\begin{equation}
\begin{split}
    \mcM_{\text{SM},\pm}&=\frac{i g_2^2}{2}\,\frac{\mw}{ \sqrt{\hat{s}}}(1\mp\cos\theta) \:\frac{1}{1-\varepsilon_W^2}\\
    \mcM_{\text{SM},0}&=\frac{i g_2^2}{2\sqrt{2}}\,\sin\theta\:\frac{1+\varepsilon_W^2-\varepsilon_H^2 }{1-\varepsilon_W^2}
\end{split}
\end{equation}

\begin{equation}
\begin{split}
    \mcM_{\varphi q,\pm}^{(3)} &=
    2 i \,c_{\varphi q}^{(3)} \frac{\sqrt{\hat{s}}\mw}{\Lambda^2}\,(1\mp\cos\theta)\:\frac{1}{1-\varepsilon_W^2}\\
    \mcM_{\varphi q,0}^{(3)} &= i \sqrt{2}\,c_{\varphi q}^{(3)} \frac{\hat{s}}{\Lambda^2} \,\sin\theta
   \:\frac{1+\varepsilon_W^2-\varepsilon_H^2 }{1-\varepsilon_W^2} 
\end{split}
\end{equation}

\begin{equation}
\begin{split}
    \mcM_{\varphi {\textsc w},\pm} &= 2 i\, c_{\varphi {\textsc w}} \frac{\sqrt{\hat{s}}\mw}{\Lambda^2} \,(1\mp\cos\theta)
        \:\frac{1+\varepsilon_W^2-\varepsilon_H^2 }{1-\varepsilon_W^2} \\
    \mcM_{\varphi {\textsc w},0} &=4 i \sqrt{2}\,c_{\varphi  {\textsc w}}\frac{\mw^2}{\Lambda^2}\, \sin\theta \:\frac{1}{1-\varepsilon_W^2} \\
\end{split}
\end{equation}

\begin{equation}
\begin{split}
    \mcM_{\varphi \widetilde{\textsc w},\pm} &= 2 \, c_{\varphi \widetilde{\textsc w}} \frac{\sqrt{\hat{s}}\mw}{\Lambda^2} \,(1\mp\cos\theta)
        \:\frac{\lambda(\varepsilon_W,\varepsilon_H)}{1-\varepsilon_W^2} \\
    \mcM_{\varphi \widetilde{\textsc w},0} &= 0\,,
\end{split}
\end{equation}
where $\lambda(\varepsilon_W,\varepsilon_H)\equiv\sqrt{(1+\varepsilon_W+\varepsilon_H)(1+\varepsilon_W-\varepsilon_H)(1-\varepsilon_W+\varepsilon_H)(1-\varepsilon_W-\varepsilon_H)}$ is sometimes referred to as the triangle function.

\subsection{$p p \to Wh \to \ell \nu h$}
\label{app:amps_squared}
\newcommand{\dwsq}{\left|\mathcal{D}_W\right|^2}
Here, we write the full squared amplitudes for $Wh$ production with $W\to\ell\nu$. The expressions are expanded in $M_W/\sqrt{\hat{s}}$ up to the order where even functions of $\phi_W$ appear in order to capture the dependence on $\phi_W$ in the presence of the neutrino momentum reconstruction
ambiguity discussed in section~\ref{sec:interference}.

The squared amplitudes and the interference terms between one BSM amplitude and the SM are given separately. The three BSM-BSM interference amplitudes are omitted for brevity since we are mainly interested in the regime where the dependence on the Wilson coefficients is linear. For convenience, we define the square of the $W$  propagator denominator as $\dwsq\equiv(M_{\ell\nu}^2-M_W^2)^2+M_W^2\Gamma_W^2$. 
Note that the angular dependence on the $W$ scattering angle, $\theta$, in Eq.~\ref{eq:inclusiveInterf} can be recovered by integrating over the phase space of the leptons, $d\cos\theta_W d\phi_W$.

\begin{multline}
    \left|\mathcal{M}_\text{SM}\right|^2 = 
    \frac{g_2^6}{48}\Bigg[
    \frac{1}{4} \sin^2\theta \sin^2\theta_W+
    \frac{M_W}{\sqrt{\hat{s}}} \left(1-\cos\theta\cos\theta_W\right)
	    \sin\theta  \sin\theta_W \cos\phi_W\\+
	\frac{M_W^2}{\hat{s}}\left(1-\cos\theta \cos \theta_W\right){}^2\cos^2\phi_W
	\Bigg]\frac{M_W^2}{\dwsq}
\end{multline}
\begin{multline}
2\Re\mathcal{M}_\text{SM}\mathcal{M}_{\varphi q}^{(3)*} =   \frac{c_{\varphi q}^{(3)} g_2^4}{6}\frac{\hat{s}}{\Lambda^2}\Bigg[
\frac{1}{4}\sin^2\theta \sin^2\theta_W+
\frac{M_W}{\sqrt{\hat{s}}} \left(1-\cos\theta \cos \theta_W\right)\sin\theta\sin\theta_W \cos \phi_W\\
  +\frac{M_W^2}{\hat s} \left\{
    \left(1-\cos\theta \cos\theta_W\right)^2+\sin^2\theta\sin^2\theta_W \cos ^2\phi_W
 \right\}
  -\frac{M_H^2}{2\hat s} \sin^2\theta\sin^2\theta_W
\Bigg]\frac{M_W^2}{\dwsq}
\end{multline}
\begin{multline}
2\Re\mathcal{M}_\text{SM}\mathcal{M}_{\varphi {\textsc w}}^* = 
-\frac{g_2^4 c_{\varphi {\textsc w}}}{6} \frac{M_W^2}{\Lambda^2} \Bigg[
\left(1-\cos\theta\cos\theta_W\right)^2 +
\sin^2\theta\sin^2\theta_W\cos^2\phi_W \\
+ \frac{1}{2}\frac{\sqrt{\hat{s}}}{M_W}  \left(1-\cos\theta \cos\theta_W\right)\sin\theta \sin\theta_W \cos\phi_W
\Bigg]\frac{M_W^2}{\dwsq}
\end{multline}
\begin{multline}
2\Re\mathcal{M}_\text{SM}\mathcal{M}_{\varphi \widetilde{\textsc w}}^* = 
\frac{g_2^4 c_{\varphi \widetilde{\textsc w}}}{12} \frac{\sqrt{\hat s}\mw}{\Lambda^2} \Bigg[
 \left(1-\cos\theta \cos\theta_W\right)\sin\theta \sin\theta_W \sin\phi_W
\Bigg]\frac{M_W^2}{\dwsq}
\end{multline}
\begin{multline}
\left|\mathcal{M}_{\varphi q}^{(3)}\right|^2 = 
\frac{g_2^2 c_{\varphi q}^{(3)2}}{3}\frac{\hat{s}^2}{\Lambda^4}\Bigg[
\frac{1}{4}\sin^2\theta\sin^2\theta_W
+\frac{M_W}{\sqrt{\hat s}} \left(1-\cos\theta\cos \theta_W\right)\sin\theta \sin\theta_W\cos\phi_W\\
 +\frac{M_W^2}{\hat{s}} \left(\sin^2\theta\sin^2\theta_W\cos^2\phi_W +\left(\cos\theta-\cos\theta_W\right)^2\right)
\Bigg]\frac{M_W^2}{\dwsq}
\end{multline}
\begin{multline}
\left|\mathcal{M}_{\varphi {\textsc w}}\right|^2 =\frac{g_2^2 \, c_{\varphi  {\textsc w}}^2}{3} \frac{\hat{s} \mw^2}{\Lambda^4} \Bigg[
\sin^2\theta \sin^2\theta_W \cos ^2\phi_W+\left(\cos
\theta-\cos\theta_W\right)^2\\
+\frac{M_W }{\sqrt{\hat{s}}} \sin\theta \sin\theta_W \left(1-\cos\theta \cos
\theta_W\right) \cos \left(\phi_W\right)
\Bigg]\frac{M_W^2}{\dwsq}
\end{multline}
\begin{multline}
\left|\mathcal{M}_{\varphi \widetilde{\textsc w}}\right|^2 =\frac{g_2^2\,c_{\varphi  \widetilde{\textsc w}}^2}{3}  \frac{\hat{s}\mw^2}{\Lambda^4} \Bigg[
\left(1-\cos\theta \cos\theta_W\right)^2-\sin^2\theta\sin^2\theta_W\cos^2\phi_W
\Bigg]\frac{M_W^2}{\dwsq}
\end{multline}

\section{Monte Carlo event generation}\label{app:mc_evt_gen}
In this appendix, we provide some additional details regarding the Monte Carlo event generation discussed in Section~\ref{sec:signal_background}. First, as mentioned above, we take the rate for a jet to fake a photon to be $P_{j\to\gamma}=10^{-3}$. This rate is conservative with respect to the fake rates reported in~\cite{Contino:2016spe,Abada:2019lih}, which are parametrized as $P_{j\to\gamma}= 0.01 \, \exp(-p_T^\gamma/(30 \text{GeV}))$ and $0.002 \, \exp(-p_T^\gamma/(30 \text{GeV}))$, respectively.
Even with this conservative fake rate, the $Wjj$ background is subleading while the $W\gamma j$ one is of the same order as $W\gamma\gamma$. Nevertheless, their sum is still much smaller than the signal (see Fig.~\ref{fig.pTh}), hence we do not expect their reduction to have a significant effect on the bound.

As for the event samples themselves, processes without a jet in the final state were generated inclusively with one additional hard jet. The 0- and 1-jet samples were matched in the MLM scheme as implemented in \textsc{MadGraph}. The $k_\perp-$cutoff scale was set to $1/3\cdot p_{T,\min\{\text{bin}\}}^{h/\gamma\gamma}$ when generating the background ($1/2\cdot p_{T,\min\{\text{bin}\}}^{h/\gamma\gamma}$ for the signal), where $\min\{\text{bin}\}$ is the lower edge of the generation bin.
The main reason for this was to account for new production channels with initial gluons. On the other hand, the backgrounds with at least one jet at generation level already have those channels open without the need for an extra hard jet.

Fully differential NNLO QCD corrections to $Wh$ production were obtained in~\cite{Ferrera:2011bk,Ferrera:2013yga,Campbell:2016jau,Ferrera:2017zex} including mass effects in the decays and matched to a parton shower in~\cite{Astill:2016hpa}. Generically, the NNLO/NLO $k$-factors as a function of $\pth$ are small ($<10\%)$. The same is also true of the NLO/LO $k$-factors if the LO is showered.
In our case, with the $0+1j$ matched sample, the NLO/$0+1j$ $k$-factor is $25-50\%$ but this difference comes mainly from the choice of PDFs. As mentioned above, we used \texttt{NNPDF23LO} for the $0+1j$ sample. However, at NLO, we used the \texttt{NNPDF23NLO} PDFs for consistency. The NLO/$0+1j$ $k$-factor becomes $\leq 10\%$ if one generates the $0+1j$ sample using the NLO PDFs.

The inclusive electroweak (EW) corrections to this process were computed in~\cite{Ciccolini:2003jy} and the fully-differential corrections in~\cite{Denner:2011id,Granata:2017iod}.
They were included in \verb|MG5_aMC@NLO| in~\cite{Frederix:2018nkq}.
While EW corrections are known to be large for large $\pth$, their effect on our analysis is $\lesssim 20\%$;
nevertheless, we applied the $k$-factors extracted from~\cite{Frederix:2018nkq} to the signal process.
The recomputed $k$-factors in our first four $\pth$ bins defined in Eq.~\eqref{eq:binning} are $\{0.92,0.85,0.79,0.73\}$ while we applied an estimated $k$-factor of 0.6 in the overflow ($\pth>1$ TeV) bin. Note that the overflow bin does not contribute to the bound and therefore does not warrant a more careful estimate.

\subsection{Generation cuts}

\begin{table}[t]
\centering{
\renewcommand{\arraystretch}{1.25}
\begin{tabular}{ c @{\hspace{.5em}} | @{\hspace{.5em}} c  @{\hspace{1.em}} c   @{\hspace{1.em}} c  }
\toprule
& $Wh$ & $W\gamma\gamma$ & \hspace{0.1cm} $Wj\gamma$ and $Wjj$ \\\midrule
$p_{T,\min}^\ell$ [GeV] & \multicolumn{3}{c}{30\hspace{1.5em} \small (all samples)}\\
$p_{T,\min}^{\gamma,j}$ [GeV] & \multicolumn{3}{c}{50\hspace{1.5em} \small (all samples)}\\
$\slashed{E}_{T,\min}$ [GeV] & \multicolumn{3}{c}{100\hspace{1.5em} \small (all samples)}\\
$|\eta^{j,\ell}_{\max}|$ & \multicolumn{3}{c}{6.1\hspace{1.5em} \small (all samples)}\\\midrule\midrule
$\Delta R^{\gamma\gamma,\gamma j,\gamma\ell}_{\min}$ & -- & 0.01 & 0.01\\
$\Delta R^{\gamma\gamma,\gamma j,jj}_{\max}$ & -- & 2.5 & 2\\
$m^{\gamma\gamma, \gamma j, jj}$ [GeV] & -- & [50,300] & [50,250] \\
$p_{T,\min}^{h,\gamma\gamma}$ [GeV]& \{150,350,550,750\} & \{100,300,500,700\} & --\\
$p_{T,\min}^{\ell\nu}$ [GeV]& -- & -- &  \{100,300,500,700\}\\
\bottomrule
\end{tabular}}
\caption{Parton level generation cuts 
    for the signal and background processes.
    Each element in the list of values for $p_{T}^{h,\gamma\gamma}$ and $p_{T}^{\ell\nu}$ corresponds to the  cut used in 4 different generation runs. Each run was used in the analysis of the corresponding $p_{T}^{h}$ bin.
    The last generation bin is used for both the fourth and fifth (overflow) bin.
}
\label{tab:gen_cuts}
\end{table}
\begin{table}[t]
   \centering
   \begin{tabular}{c|>{\centering\arraybackslash}p{0.1\textwidth}|>{\centering\arraybackslash}p{0.1\textwidth}|>{\centering\arraybackslash}p{0.1\textwidth}|>{\centering\arraybackslash}p{0.1\textwidth}}
    \toprule
      & $h \to \gamma \gamma$ & $\gamma \gamma$ & $j \gamma$ & $j j$ \\
         \hline
     $\sigma_\text{(loose)}$ [fb] & $7.5$ & $4.8\cdot 10^3$ & $10^6$ & $6.2\cdot 10^7$  \\
     \hline
      $\sigma_{\text{(} 3^{\text{rd}} \text{ bin gen. cuts)}}$ [fb] & $0.026$ & $2.9$ &$3.0 \cdot 10^2$ & $5.2 \cdot 10^3$  \\
   \bottomrule
     \end{tabular}
\caption{Parton level cross sections for signal and backgrounds before and after imposing the generation level cuts defined in Table \ref{tab:gen_cuts}. Signal and $\gamma \gamma$ were generated at (0+1$j$), while $j\gamma$ and $jj$ are LO. We only show the cross section after generation cuts for the third bin.
See text for more details. The subscript `loose' refers to the mild cuts we had to impose to regulate infrared divergencies. We employed the cuts $p_T^{j,\gamma}>20\,\text{GeV}$ for all the four processes, $m_X>20\,\text{GeV}$ for the process $pp\rightarrow l\nu X$,
$\Delta R_\text{min}^{\gamma j, \gamma l}=0.01$ for $X=\gamma \gamma$ and $\Delta R_\text{min}^{\gamma l}=0.01$ for $X=\gamma j$. }
\label{tab:gen_bin3XS}
\end{table}

In order to have more Monte Carlo events after the selection cuts, we imposed several basic cuts at generation level which we list in Table~\ref{tab:gen_cuts}. On the upper part of Table~\ref{tab:gen_cuts}, we show the cuts common to all the channels and all bins. On the lower part of the table, we show the cuts corresponding to each channel and each bin. The bin-specific cuts are done in order to increase the number of Monte Carlo events falling in each $p_T^h$ bin defined in Eq.~\eqref{eq:binning}, without cutting any events that could pass the detector simulation and subsequent selection cuts. The generation cuts are not one to one with the $p_T^h$ bins, because this quantity is shifted due to showering.

For illustration, we show in Table~\ref{tab:gen_bin3XS} the cross section before and after the generation cuts described in Table~\ref{tab:gen_cuts}. After generation cuts, we only give the results for the events in the third bin, since it is the most sensitive one as a showcase example.
Notice that the generation cuts are slightly different for each process, therefore the interpretation of the relative size of the cross sections before and after generation cuts
must be taken with care.

\newpage
\section{Fits of the signal cross section}\label{app:other}
In Table \ref{tab:sigma_full}, we show the fits of the $Wh \to \ell \nu \gamma \gamma$ cross section as a function of the $c_{\varphi q}^{(3)}$, $c_{\varphi {\textsc w}}$ and $c_{\varphi \widetilde {\textsc w}}$ Wilson coefficients for the bins used in the global analysis of Section~\ref{sec:global_fit}.

\begin{table}[h!!]
\begin{centering}
\setlength{\extrarowheight}{0mm}%
\scalebox{.86}{
\begin{tabular}{c|c|c|c}
\toprule
\rule[-.5em]{0pt}{.5em}
\multirow{2}{*}{$p_T^{h}$ bin} &\multirow{2}{*}{$\phi_{W}$ bin}& \multicolumn{2}{c}{Number of expected events}\tabularnewline
\cline{3-4} &  & Signal & Background \tabularnewline
\hline 
\multirow{2}{*}{\rule{0pt}{3.5em}$[200-400]$\,GeV} & $[-\pi,0]$ &
$\begin{aligned}\rule{0pt}{1.15em}1310 & + 10380\,c_{\varphi q}^{(3)}
+ 1290\,c_{\varphi {\textsc w}} + 641\,c_{\varphi\widetilde {\textsc w}}\\
&+ 25700\,\big(c_{\varphi q}^{(3)}\big)^{2}
+ 1510\,\left(c_{\varphi {\textsc w}}\right)^{2}
+1350\,\big(c_{\varphi\widetilde {\textsc w}}\big)^{2}\\
&+ 5912\,c_{\varphi q}^{(3)}\,c_{\varphi {\textsc w}}
+3402\,c_{\varphi\widetilde {\textsc w}}\,c_{\varphi q}^{(3)}
+ 234\,c_{\varphi {\textsc w}}\,c_{\varphi\widetilde {\textsc w}}\rule[-.5em]{0pt}{1.em}
\end{aligned}
$ & $830$\tabularnewline
\cline{2-4} 
 & $[0,\pi]$ & $\begin{aligned}\rule{0pt}{1.15em}1310\, & + 10480\,c_{\varphi q}^{(3)}
+ 1250\,c_{\varphi {\textsc w}}
- 651\,c_{\varphi\widetilde {\textsc w}}\\
&+ 27000\,\big(c_{\varphi q}^{(3)}\big)^{2}
+ 1470\,\left(c_{\varphi {\textsc w}}\right)^{2}
 + 1400\,\big(c_{\varphi\widetilde {\textsc w}}\big)^{2}\\
&+ 5770\,c_{\varphi q}^{(3)}\,c_{\varphi {\textsc w}}
- 2390\,c_{\varphi\widetilde {\textsc w}}\,c_{\varphi q}^{(3)}
- 153\,c_{\varphi {\textsc w}}\,c_{\varphi\widetilde {\textsc w}}\rule[-.5em]{0pt}{1.em}
\end{aligned}
$ & $960$\tabularnewline
\hline 
\multirow{2}{*}{\rule{0pt}{3.5em}$[400-600]$\,GeV} & $[-\pi,0]$ & $\begin{aligned}\rule{0pt}{1.15em}284\, & + 5820\,c_{\varphi q}^{(3)}
+288\,c_{\varphi {\textsc w}}
+ 262\,c_{\varphi\widetilde {\textsc w}}\\
& +35800\,\big(c_{\varphi q}^{(3)}\big)^{2}
+872\,\left(c_{\varphi {\textsc w}}\right)^{2}
+834\,\big(c_{\varphi\widetilde {\textsc w}}\big)^{2}\\
& +2900\,c_{\varphi q}^{(3)}\,c_{\varphi {\textsc w}}
+3400\,c_{\varphi\widetilde {\textsc w}}\,c_{\varphi q}^{(3)}
+72.3\,c_{\varphi {\textsc w}}\,c_{\varphi\widetilde {\textsc w}}\rule[-.5em]{0pt}{1.em}
\end{aligned}
$ & $119$\tabularnewline
\cline{2-4} 
 & $[0,\pi]$ & $\begin{aligned}\rule{0pt}{1.15em}283\, & + 5860\,c_{\varphi q}^{(3)}
+ 287\,c_{\varphi {\textsc w}}
- 255\,c_{\varphi\widetilde {\textsc w}}\\
&   +36000\,\big(c_{\varphi q}^{(3)}\big)^{2}
+876\,\left(c_{\varphi {\textsc w}}\right)^{2}
+835\,\big(c_{\varphi\widetilde {\textsc w}}\big)^{2}\\
&+3260\,c_{\varphi q}^{(3)}\,c_{\varphi {\textsc w}}
-3760 \,c_{\varphi\widetilde {\textsc w}}\,c_{\varphi q}^{(3)}
-75.7 \,c_{\varphi {\textsc w}}\,c_{\varphi\widetilde {\textsc w}}\rule[-.5em]{0pt}{1.em}
\end{aligned}
$ & $129$\tabularnewline
\hline 
\multirow{2}{*}{\rule{0pt}{3.5em}$[600-800]$\,GeV} & $[-\pi,0]$ & $\begin{aligned}\rule{0pt}{1.15em}70\, & + 2760\,c_{\varphi q}^{(3)}
+ 69.4\,c_{\varphi {\textsc w}}
+ 98.9\,c_{\varphi\widetilde {\textsc w}}\\
&+33500\,\big(c_{\varphi q}^{(3)}\big)^{2}
+446\,\left(c_{\varphi {\textsc w}}\right)^{2}
+439\,\big(c_{\varphi\widetilde {\textsc w}}\big)^{2}\\
&+1830\,c_{\varphi q}^{(3)}\,c_{\varphi {\textsc w}}
+2660\,c_{\varphi\widetilde {\textsc w}}\,c_{\varphi q}^{(3)}
+28.2\,c_{\varphi {\textsc w}}\,c_{\varphi\widetilde {\textsc w}}\rule[-.5em]{0pt}{1.em}
\end{aligned}
$ & $21$\tabularnewline
\cline{2-4} 
 & $[0,\pi]$ & $\begin{aligned}\rule{0pt}{1.15em}70\, &+ 2850\,c_{\varphi q}^{(3)}
+ 74.1\,c_{\varphi {\textsc w}}
- 102\,c_{\varphi\widetilde {\textsc w}}\\
&  +33800\,\big(c_{\varphi q}^{(3)}\big)^{2}
+452\,\left(c_{\varphi {\textsc w}}\right)^{2}
+427\,\big(c_{\varphi\widetilde {\textsc w}}\big)^{2}\\
&+1380\,c_{\varphi q}^{(3)}\,c_{\varphi {\textsc w}}
-2520\,c_{\varphi\widetilde {\textsc w}}\,c_{\varphi q}^{(3)}
-24.3\,c_{\varphi {\textsc w}}\,c_{\varphi\widetilde {\textsc w}}\rule[-.5em]{0pt}{1.em}
\end{aligned}
$ & $22$ \tabularnewline
\hline 
\multirow{2}{*}{\rule{0pt}{3.5em}$[800-1000]$\,GeV} & $[-\pi,0]$ & $\begin{aligned}\rule{0pt}{1.15em}15\, & + 947\,c_{\varphi q}^{(3)}
+ 15.2\,c_{\varphi {\textsc w}}
+ 27.8\,c_{\varphi\widetilde {\textsc w}}\\
&+17900\,\big(c_{\varphi q}^{(3)}\big)^{2}
+159\,\left(c_{\varphi {\textsc w}}\right)^{2}
+147\,\big(c_{\varphi\widetilde {\textsc w}}\big)^{2}\\
&+653\,c_{\varphi q}^{(3)}\,c_{\varphi {\textsc w}}
+864\,c_{\varphi\widetilde {\textsc w}}\,c_{\varphi q}^{(3)}
+5.54\,c_{\varphi {\textsc w}}\,c_{\varphi\widetilde {\textsc w}}\rule[-.5em]{0pt}{1.em}
\end{aligned}
$ & $3$ \tabularnewline
\cline{2-4} 
 & $[0,\pi]$ & $\begin{aligned}\rule{0pt}{1.15em}15\, & + 947\,c_{\varphi q}^{(3)}
+ 15.3\,c_{\varphi {\textsc w}}
- 28.8\,c_{\varphi\widetilde {\textsc w}}\\
& +18200\,\big(c_{\varphi q}^{(3)}\big)^{2}
+156\,\left(c_{\varphi {\textsc w}}\right)^{2}
 +149\,\big(c_{\varphi\widetilde {\textsc w}}\big)^{2}\\
& +541\,c_{\varphi q}^{(3)}\,c_{\varphi {\textsc w}}
-1150\,c_{\varphi\widetilde {\textsc w}}\,c_{\varphi q}^{(3)}
-10.3\,c_{\varphi {\textsc w}}\,c_{\varphi\widetilde {\textsc w}}\rule[-.5em]{0pt}{1.em}
\end{aligned}
$ & $5$ \tabularnewline
\hline 
\multirow{2}{*}{\rule{0pt}{3.5em}$[1000-\infty]$\,GeV} & $[-\pi,0]$ & $\begin{aligned}\rule{0pt}{1.15em}4\, & + 426\,c_{\varphi q}^{(3)}
+ 4.12\,c_{\varphi {\textsc w}}
+ 9.72\,c_{\varphi\widetilde {\textsc w}}\\
&+16400\,\big(c_{\varphi q}^{(3)}\big)^{2}
+73.2\,\left(c_{\varphi {\textsc w}}\right)^{2}
+69.7\,\big(c_{\varphi\widetilde {\textsc w}}\big)^{2}\\
&+281\,c_{\varphi q}^{(3)}\,c_{\varphi {\textsc w}}
+955\,c_{\varphi\widetilde {\textsc w}}\,c_{\varphi q}^{(3)}
+1.56\,c_{\varphi {\textsc w}}\,c_{\varphi\widetilde {\textsc w}}\rule[-.5em]{0pt}{1.em}
\end{aligned}
$ & $2$ \tabularnewline
\cline{2-4} 
 & $[0,\pi]$ & $\begin{aligned}\rule{0pt}{1.15em}4\, & + 428\,c_{\varphi q}^{(3)}
+ 4.23\,c_{\varphi {\textsc w}}
- 10.6\,c_{\varphi\widetilde {\textsc w}}\\
& +16600\,\big(c_{\varphi q}^{(3)}\big)^{2}
+71.4\,\left(c_{\varphi {\textsc w}}\right)^{2}
 +69.7\,\big(c_{\varphi\widetilde{\textsc w}}\big)^{2}\\
& +226\,c_{\varphi q}^{(3)}\,c_{\varphi {\textsc w}}
-740\,c_{\varphi\widetilde {\textsc w}}\,c_{\varphi q}^{(3)}
-3.13\,c_{\varphi {\textsc w}}\,c_{\varphi\widetilde {\textsc w}}\rule[-.5em]{0pt}{1.em}
\end{aligned}$
 & $1$ \tabularnewline
\bottomrule
\end{tabular}
}
\par\end{centering}
\caption{Number of expected signal and background events at FCC-hh with $30\,{\rm ab}^{-1}$. For the signal, it is given as a function of the Wilson coefficients (with $\Lambda = 1 \, \text{TeV}$).  Notice that the coefficients have errors of order $\it few$ percent due to statistical fluctuations. The contribution of $Wjj$ to the background events is neglected.
}
\label{tab:sigma_full}
\end{table}

\clearpage
\providecommand{\href}[2]{#2}\begingroup\raggedright\endgroup

\end{document}